\documentclass[structabstract]{aa}  

\usepackage{graphicx}
\usepackage{natbib}
\usepackage{amsmath}
\usepackage{txfonts}
\def\SeBa{SeBa}

\newcommand{\maa}{\alpha \alpha} 
\newcommand{\mga}{\gamma \alpha}
\newcommand{\thickhline}{\noalign{\hrule height 0.8pt}}
 
\def\Rsolar{$R_{\odot}$}
\def\Msolar{$M_{\odot}$}

\newcommand{\peryr}{\,{\rm yr^{-1}}}
\newcommand{\yr}{{\,\rm yr}}
\newcommand{\Mo}{M_{\odot}}
\newcommand{\Ro}{R_{\odot}}
\newcommand{\persec}{\,{\rm s^{-1}}}
\begin{document}

   \title{Supernova Type Ia progenitors from merging double white dwarfs:   }

   \subtitle{Using a new population synthesis model}

   \author{S. Toonen
          \inst{1}\fnmsep\thanks{e-mail: s.toonen@astro.ru.nl}
          \and
          G. Nelemans\inst{1,2}
          \and
          S. Portegies Zwart \inst{3} 
          }

   \institute{Department of Astrophysics, IMAPP, Radboud University Nijmegen, P.O. Box 9010, 6500 GL Nijmegen, The Netherlands 
         \and Instituut voor Sterrenkunde, KU Leuven, Celestijnenstraat 200D, 3001 Leuven, Belgium
         \and Leiden Observatory, Leiden University,  P.O. Box 9513, 2300 RA
Leiden, The Netherlands\\
             }

   \date{Received February 6, 2012; July 20, 2012 }

  \abstract
 {The study of Type Ia supernovae (SNIa) has lead to greatly improved insights into many fields in astrophysics, e.g. cosmology, and also into the metal enrichment of the universe. Although a theoretical explanation of the origin of these events is still lacking, there is a general consensus that SNIa are caused by the thermonuclear explosions of carbon/oxygen white dwarfs with masses near the Chandrasekhar mass. 
 }
 {We investigate the potential contribution to the supernova Type Ia rate from the population of merging double carbon-oxygen white dwarfs. We aim to develope a model that fits the observed SNIa progenitors as well as the observed close double white dwarf population. We differentiate between two scenarios for the common envelope (CE) evolution; the $\alpha$-formalism based on the energy equation and the $\gamma$-formalism that is based on the angular momentum equation. In one model we apply the $\alpha$-formalism always. In the second model the $\gamma$-formalism is applied, unless the binary contains a compact object or the CE is triggered by a tidal instability for which the $\alpha$-formalism is used. 
 }
 {The binary population synthesis code SeBa was used to evolve binary systems from the zero-age main sequence to the formation of double white dwarfs and subsequent mergers. SeBa has been thoroughly updated since the last publication of the content of the code. 
 }
 {
The limited sample of observed double white dwarfs is better represented by the simulated population using the $\gamma$-formalism for the first CE phase than the $\alpha$-formalism.
For both CE formalisms, we find that although the morphology of the simulated delay time distribution matches that of the observations within the errors, the normalisation and time-integrated rate per stellar mass are a factor $\sim 7-12$ lower than observed. 
Furthermore, the characteristics of the simulated populations of merging double carbon-oxygen white dwarfs are discussed and put in the context of alternative SNIa models for merging double white dwarfs. 
 }
 {}
   \maketitle

%________________________________________________________________

\section{Introduction}
Type Ia supernovae (SNIa) are one of the most energetic explosive events known. 
They have been of great importance in many fields, most notably as a tool in observational cosmology. They have been used very successfully as standard candles on cosmological distance scales \citep[e.g.][]{Rie98, Per99}, owing to the special property of great uniformity in the light curves \citep[e.g.][]{Phi93}. 
The SNIa also strongly affect the Galactic chemical evolution through the expulsion of iron \cite[e.g.][]{Ber91}.  Despite their significance Type Ia supernovae are still poorly understood theoretically. 

Supernovae Type Ia are generally thought to be caused by thermonuclear explosions of carbon/oxygen (CO) white dwarfs (WDs) with masses near the Chandrasekhar mass $M_{\rm ch}\approx 1.4\Mo$ \citep[e.g.][]{Nom82}. Various progenitor scenarios have been proposed. The standard scenarios can be divided into two schools of thoughts: the single-degenerate (SD) \citep{Whe73} and double-degenerate (DD) scenario \citep{Web84, Ibe84}. In the SD scenario, a CO WD explodes as an SNIa if its mass approaches $M_{\rm ch}$ through accretion from a non-degenerate companion. In the DD scenario, two CO WDs can produce an SN Ia while merging if their combined mass is larger than $M_{\rm ch}$. 

However, observationally as well as theoretically, the exact nature of the SNIa progenitors remains unclear. 
The explosion mechanism is complex due to the interaction of hydrodynamics and nuclear reactions. Several models exist that vary for example between a detonation or deflagration disruption or vary between explosions at the Chandrasekhar mass or sub-Chandrasekhar masses \citep[see e.g.][for a review]{Hil00}. 
It also remains unclear whether the DD and SD scenario both contribute to the SNIa rate or if one of the scenarios dominates.
Both scenarios have problems in matching theories with observations. A serious concern about the DD scenario is whether the collapse of the remnant would lead to a supernova or to a neutron star through accretion-induced collapse \citep[see][]{Nom85, Sai85, Pie03, Yoo07, Pak10, Pak12, She12}. Although in the SD channel the models for the explosion process need to be fine-tuned to reproduce the observed spectra and light curves, an SNIa like event is more easily reproduced in the simulations of the explosion process. One problem with the SD scenario is that the white dwarfs should go through a long phase of supersoft X-ray emission, although it is unclear if there are enough of these sources to account for the SNIa rate \citep[see][]{DiS10, Gil10, Hac10}. Moreover archival data of known SNIa have not shown this emission unambiguously, but there is may be one case \citep[see][]{Vos08, Roe08, Nie10}. 
Furthermore, SNIa that take place more than a few 10$^9$ years after the starburst \citep[see e.g.][]{Mao10} are hard to create in this channel \citep[e.g.][]{Yun00, Han04}. 
%it is hard to create SNIa in this channel at times more than a few 10$^9$ years after starburst \citep[e.g.][]{Yun00, Han04} that have been observed \citep[e.g.][]{Mao10}. 

To use SNIa as proper standard candles, we need to know what SNIa are, when they happen and what their progenitors are. Therefore, we study the binary evolution of low- and intermediate mass stars. In a forthcoming paper (Bours, Toonen \& Nelemans, in preparation) we study the SD-scenario by looking into the poorly understood physics of accretion onto white dwarfs. In this paper we focus on the DD scenario and the effect of the as yet very uncertain phases of common envelope (CE) evolution on the double white dwarf (DWD) population. These DWD systems are interesting sources for studying various phases of stellar evolution, in our case the CE evolution. Gravitational wave emission is also important as this affects the binary system by decreasing the orbital period and eventually leading to a merger \citep{Kra62, Pet64}, or possible a SNIa. The DWDs are expected to be the dominant source \citep{Eva87, Nel01c} of gravitational waves for the future space-born gravitational wave observatories such as eLISA \citep{Ama12, Ama12b}.

We study the population of merging DWDs that might lead to a SNIa from a theoretical point of view. We incorporated results from observations where possible. We use the population synthesis code \SeBa\ for simulating the stellar and binary evolution of stellar systems that leads to close DWDs. In Sect. \ref{sec:seba} we describe the code and the updates since the last publication of \SeBa. A major influence on the merging double-degenerate population is the poorly understood CE phase \citep{Pac76, Web84, Nel00}. We adopt two different models for the CE. In Sect. \ref{sec:ce} we describe these models and their implications for the observations of close DWDs. In Sect. \ref{sec:path} we discuss the binary paths leading to SNIa for each model. The SNIa rates and time-integrated numbers are derived in Sect. \ref{sec:dtd}. 
The properties of the population of merging DWDs are discussed in the context of the classical and  alternative sub- and super-Chandrasekhar SNIa explosion models in Sect. \ref{sec:pop}. A discussion and conclusion follows in Sect. \ref{sec:con}.

    \begin{figure*}
    \centering
    \begin{tabular}{c c}
	\includegraphics[angle=270, scale = 0.4]{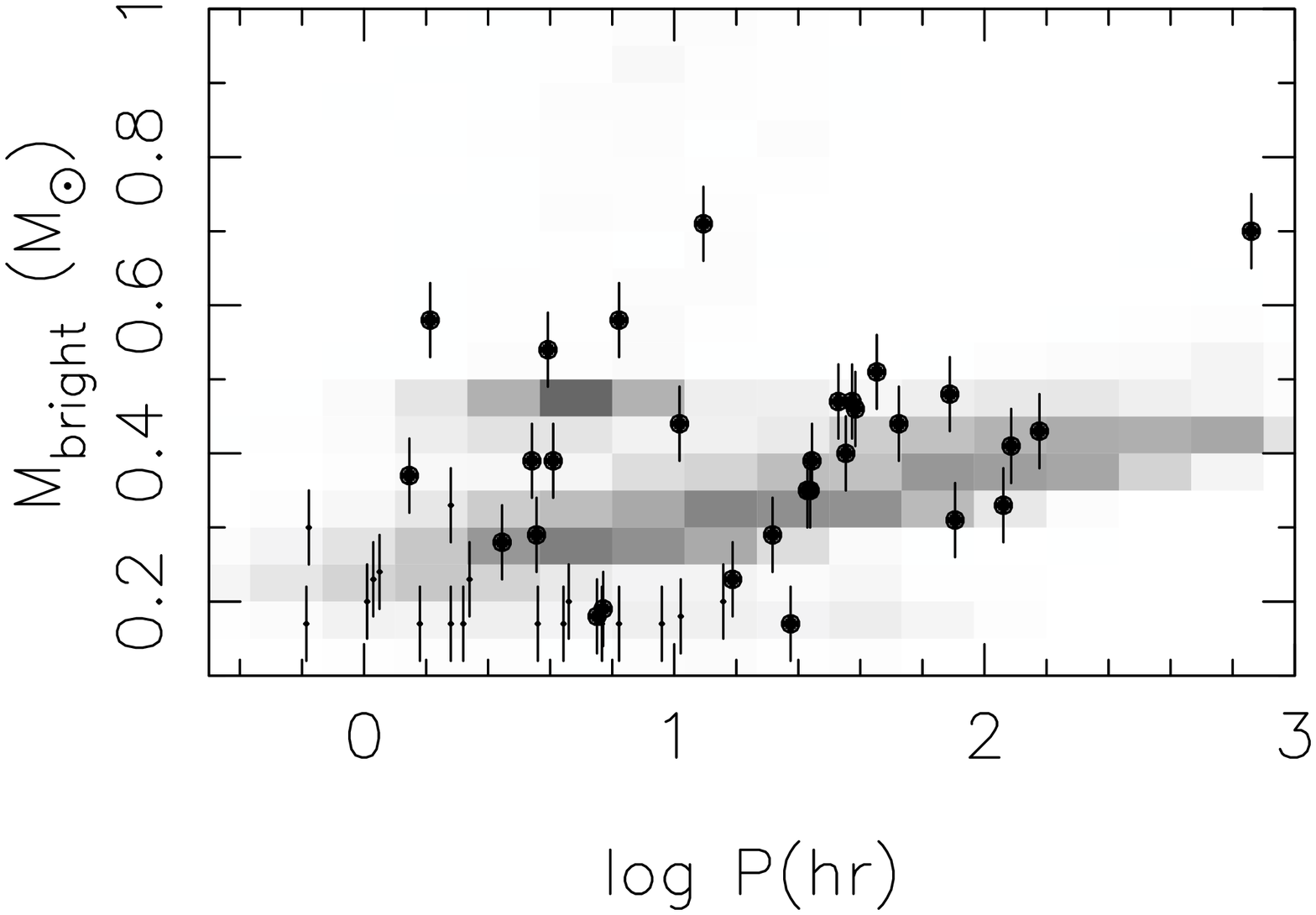} &
	\includegraphics[angle=270, scale = 0.4]{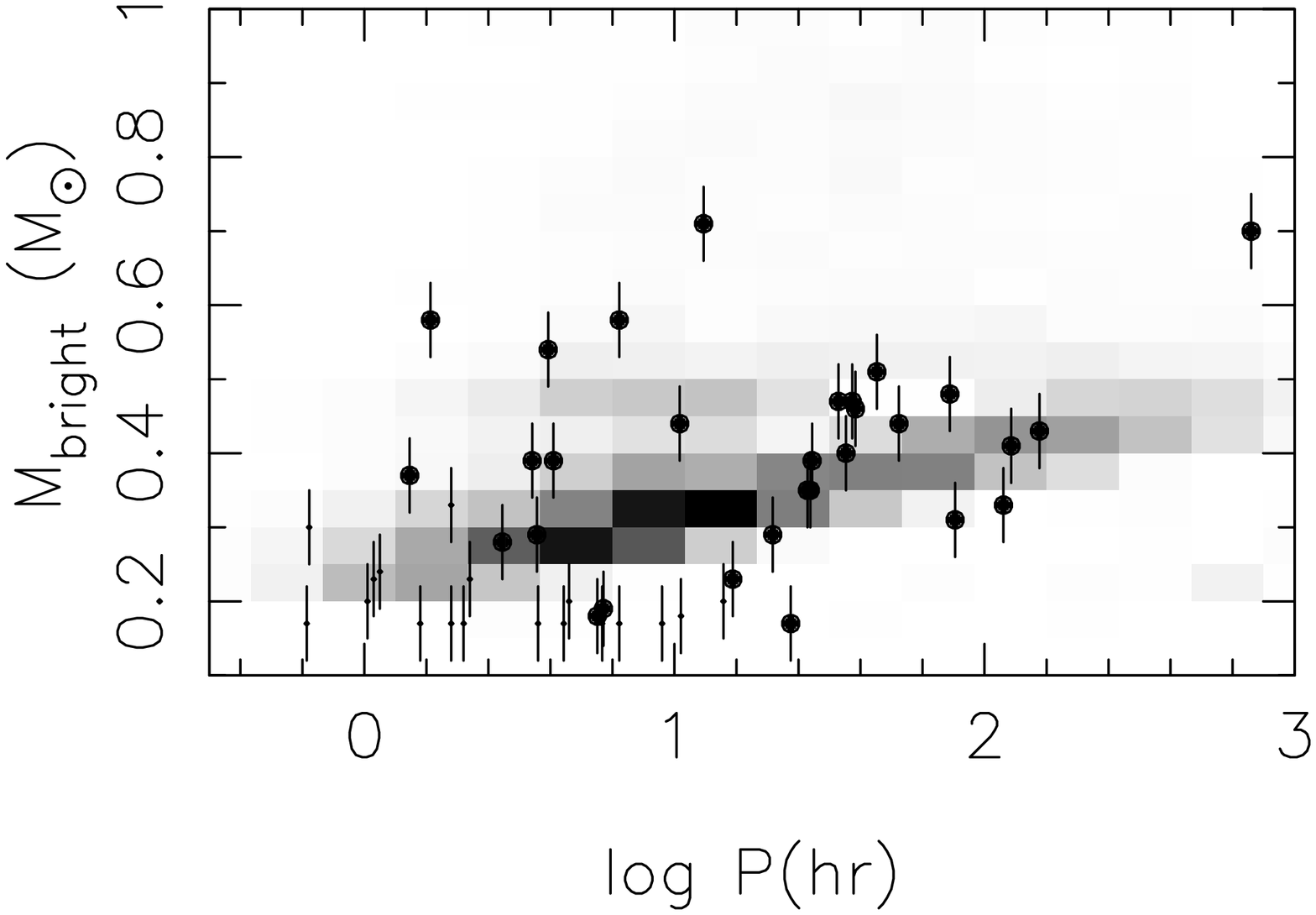} \\
	(a) & (b) \\
	\end{tabular}
    \caption{Simulated population of visible double white dwarfs as a function of orbital period and mass of the brighter white dwarf. Left: the stellar evolution tracks according to EFT are used; right: HPT (using model $\mga$, see Sect. \ref{sec:ce}). The intensity of the grey scale corresponds to the density of objects on a linear scale. The same grey scale is used for both plots. Observed binary white dwarfs are overplotted with filled circles. Thick points taken are from \citet{Mar11}, thinner points from \citet{Tov04, Nap05, Kul10, Bro10, Bro11, Mar11, Kil11, Kil11b, Kil11c}, see Sect. \ref{sec:seba_obs} for a discussion.} 
    \label{fig:pop_M1}
    \end{figure*}

\section{\SeBa\  - a fast stellar and binary evolution code}
\label{sec:seba}

We present an update to the software package \SeBa\ \citep{Por96, Nel01} for fast stellar and binary evolution computations. Stars are evolved from the zero-age main sequence (ZAMS) until remnant formation and beyond. Stars are parametrised by mass, radius, luminosity, core mass, etc. as functions of time and initial mass. Mass loss from winds, which is substantial e.g. for massive stars and post main-sequence stars, is included. 
Binary interactions such as mass loss, mass transfer, angular momentum loss, CE, magnetic braking, and gravitational radiation are taken into account with appropriate recipes at every timestep \citep{Por96, Por98}. 
Following mass transfer in a binary, the donor may turn into a helium-burning star without hydrogen envelope.
When the mass transfer leads to a merger between the binary stars, we estimate the resulting stellar product and follow the evolution further. 
Note that we do not solve the equations of stellar structure. The stellar tracks instead assume stellar models in hydrostatic equilibrium. When this is not the case, however the gas envelope surrounding the core may puff outward (see Appendix \ref{appendix:accretion} for details on the formalism). 
In our simulation the mass transfer rate is calculated from the relevant timescales (see Appendix \ref{appendix:stability}) and not from than the stellar radii. Therefore binary evolution is not critically dependent on out-of-equilibrium parameter values. 

The philosophy of SeBa is to not a priori define the binary's evolution, but rather to determine this at runtime depending on the parameters of the stellar system. 
When more sophisticated models become available of processes that influence stellar evolution, these can be included, and the effect can be studied without altering the formalism of binary interactions. 
An example is the accretion efficiency onto the accretor star during mass transfer. Instead of prescribing a specific constant percentage of the transferred matter to be accreted (and the rest to be lost from the system), the efficiency depends on the properties of the accreting star, such as the thermal timescale, the radius and the Roche lobe of the accretor (see Appendix \ref{appendix:accretion} for details). 
Another example is the stability of mass transfer. In our simulations the stability and rate of mass transfer are dependent on the reaction to mass change of the stellar radii and the corresponding Roche lobes.  The advantage of this is that the (de)stabilising effect of non-conservativeness of stable mass transfer \citep[see][]{Sob97} is taken into account automatically. There is no need to make the assumption in the stability calculation that stable mass transfer is conservative, as with methods that depend on the mass ratio \citep{Hje87, Tou97, Hur02}.

Since the last publication of the code content, many changes have been made. We briefly discuss the most important changes below, and provide more detail in Appendix \ref{appendix:}. First, the wind mass loss prescriptions that we implemented are mostly based on the recommendations by \citet{Hur00}. The specific prescriptions for different types of stars are described in Appendix \ref{appendix:wind}.  
Second, a summary of the treatment of accretion onto different stars can be found in Appendix \ref{appendix:accretion}. 
The accretion procedure previously used in \SeBa\ is complemented with a procedure for accretion from a hydrogen-poor star. We assume that for ordinary stars, helium-rich matter is accreted directly to the core of the star. 
The mass accretion process onto white dwarfs is updated with new efficiencies of mass retention on the surface of the white dwarf. For hydrogen accretion we have the option to choose between the efficiencies of \citet{Hac08} and \citet{Pri95}. Helium retention can be modelled according to \citet{Kat99} \citep[with updates from][]{Hac99} or \citet{Ibe96}. In this research we used the efficiencies of \citet{Hac08}, \citet{Kat99}, and \citet{Hac99}.
For a study of different retention efficiencies and the effect on the Supernova Type Ia rate using the new version of \SeBa, see Bours, Toonen \& Nelemans, in preparation. 
Third, the stability of mass transfer is based on the adiabatic and thermal response of the donor star to mass loss and the response of the Roche lobe. The adjustment of the Roche lobe is dependent on the mass transfer rate, which in turn sets the efficiency of accretion onto the accretor star, see Appendix \ref{appendix:stability}. 
Fourth, regarding the stellar tracks, previously, stellar evolution has been based on evolutionary tracks described by analytic formulae given by \citet[][hereafter EFT]{Egg89} with updates from \citet{Tou97} and helium star evolution as described by \citet{Por96} based on \citet{Ibe85}. In the new version, the evolution of ordinary stars and hydrogen-poor stars is based on \citet[][hereafter HPT]{Hur00}. We do not adopt the HPT tracks for remnants, instead we maintain our prescription \citep{Por96, Nel01}, which includes processes such as natal kick velocities to compact objects in supernovae explosions. 

\subsection{Impact on the population of double white dwarfs}
\label{sec:seba_obs}
Fig. \ref{fig:pop_M1} shows the visible close DWD population simulated by \SeBa . On the left a simulation is shown of the previous version of \SeBa\ that a.o. uses the EFT tracks and on the right we show the current version using the HPT tracks. 
Initial parameters are distributed according to the distributions described in Table \ref{tbl:init_param}. Primary masses are drawn from 0.96-11$\Mo$ to include all stars that evolve into a white dwarf in a Hubble time. For the mass ratio and eccentricity we cover the full range 0-1, and the orbital separation out to 10$^6 \Ro$. We assumed solar metallicity, unless specified otherwise. In the normalisation of the simulation we assumed that primary masses lie in the range 0.1-100$\Mo$. 
Our method to estimate the visible population of DWDs is described in \citet{Nel04}, in which the Galactic star formation history is based on \citet{Boi99} and WD cooling according to \cite{Han99}. We assume a magnitude limit of 21. 
In Fig. \ref{fig:pop_M1}, the observed DWDs are overplotted with filled circles. The systems are described by \citet{Mar11b} and references therein, as well as \citet{Tov04} and \citet{Rod10}. Additionally, we included 19 newly discovered DWDs from \citet{Kul10, Bro10, Bro11, Mar11, Kil11, Kil11b} and \citet{Kil11c}. These new systems are displayed with smaller circles and thinner lines to separate them from the previously found systems. We did  this because the observational biases are very different. The previously found systems were selected from a magnitude-limited sample down to 16-17 magnitude. The new systems are much fainter at about 20 magnitude. Moreover, most of the new systems are discovered as part of the ELM survey \citep{Bro10}. This survey focuses on finding extremely low-mass white dwarfs from follow-up observations of spectroscopically selected objects from the Sloan Digital Sky Survey. Therefore, the set of new systems is biased to lower masses. One should take this bias into account while comparing with the simulations and not take the combined set of observed systems as a representative sample of the DWD population. \citet{Kil11} showed in their Fig. 12 a visualisation of the population of visible DWDs simulated by \SeBa, where this selection effect has been taken into account. 

\begin{table}
\caption{Distributions of the initial binary parameter mass, mass ratio, orbital separation and eccentricity.}
\begin{tabular}{lc}
%\hline \hline
\thickhline
Parameter & Distribution\\
%\hline
\thickhline
Mass of single stars & Kroupa IMF $^{(1)}$ \\
Mass of binary primaries & Kroupa IMF $^{(1)}$  \\
Mass ratio & Flat distribution \\
Orbital separation & $N(a) da \propto a^{-1} da\ ^{(2)}$\\
Eccentricity & Thermal distribution $^{(3)} $ \\
\hline
\end{tabular}
\label{tbl:init_param}
\begin{flushleft}
\tablefoot{$^{(1)}$ \citet{Kro93}; $^{(2)}$ \citet{Abt83}; $^{(3)}$ \citet{Heg75}}
\end{flushleft}
\end{table}

The locations of the observed DWDs in Fig. \ref{fig:pop_M1} correspond reasonably well to the predictions of both models. The overall structure of the simulated populations from both models are similar. At masses of $\sim 0.5 \Mo$ and periods of $1-10$ hr, there is a very pronounced region in the plot from the EFT tracks that seems to be missing in the HPT plot. However, this is not really the case. These systems mainly consist of one helium (He) WD and one CO WD. Masses of CO WDs span a wider range of values in the HPT tracks, which distributes the pronounced region in EFT over a larger region in mass and period in HPT.

For a single burst of star formation the number of created DWDs within 13.5 Gyr and with an orbital period $P<1000$hr for the HPT and EFT stellar tracks is very similar; $6.9\cdot 10^{-3}$ per \Msolar\ of created stars for both models. The time-integrated merger rate is $2.4\cdot 10^{-3} \Mo^{-1}$ for HPT and $3.2\cdot 10^{-3} \Mo^{-1}$ for EFT. The current merger rate in the Milky Way according to the HPT and EFT stellar tracks is very similar; $1.4\cdot 10^{-2} \peryr$ for HPT and $1.2\cdot 10^{-2} \peryr$ for EFT, for which we have assumed a star formation history as in \citet{Nel04} based on \citet{Boi99}. 

Classically, the population of double He dwarfs is thought to dominate in number over the other types of close DWDs. Using the EFT tracks for a single burst of star formation, we predict a percentage of [He-He, He-CO, CO-CO] = [60\%, 17\%, 21\%] and a negligible number of DWDs containing oxygen/neon (ONe) dwarfs. For the HPT tracks, the percentage of double He dwarfs decreases to 38\%. The population consists of [He-He, He-CO, CO-CO]=[38\%, 27\%, 33\%] and 2\% CO - ONe dwarfs. The decrease in number of He WDs is caused by a difference in the stellar tracks related to helium ignition under degenerate conditions. As shown by \citet{Han02}, degenerate stars do not ignite helium at a fixed core mass, but instead the core mass at helium ignition is a decreasing function of the ZAMS mass of the star. Taking this into account, more WDs in close binaries are labelled CO WDs.

\section{Two models for common envelope evolution}
\label{sec:ce}
Close DWDs are believed to encounter at least two phases of mass transfer in which one of the stars loses its hydrogen envelope. In at least one of these phases mass transfer from the evolving more massive star to the less massive companion is dynamically unstable \citep{Pac76, Web84} which leads to a common envelope.
The core of the donor and companion spiral inward through the envelope, expelling the gaseous envelope around them. Because of the loss of significant amounts of mass and angular momentum, the CE phase plays an essential role in binary star evolution in particular the formation of short-period systems that contain a compact object. 

Despite of the importance of the CE phase and the enormous efforts of the community, all effort so far have not been successful in understanding the phenomenon. Several prescriptions for CE evolution have been proposed. The $\alpha$-formalism \citep{Web84} is based on the conservation of orbital energy. The $\alpha$-parameter describes the efficiency with which orbital energy is consumed to unbind the CE according to
\begin{equation}
E_{\rm gr} = \alpha (E_{\rm orb,init}-E_{\rm orb,final}),
\label{eq:alpha-ce}
\end{equation}
where $E_{\rm orb}$ is the orbital energy and $E_{\rm gr}$ is the binding energy between the envelope mass $M_{\rm env}$ and the mass of the donor $M$. $E_{\rm gr}$ is often approximated by
\begin{equation}
E_{\rm gr} = \frac{GM M_{\rm env}}{\lambda R},
\label{eq:Egr}
\end{equation} 
where $R$ is the radius of the donor star and $\lambda$ depends on the structure of the donor. We assume $\alpha \lambda = 2$. \citet{Nel00} deduced this value from reconstructing the last phase of mass transfer for 10 known DWDs using the unique core-mass -- radius relation for giants. 

To explain the observed distribution of DWDs, \citet{Nel00} proposed an alternative formalism. According to this $\gamma$-formalism, mass transfer is unstable and non-conservative. The mass-loss reduces the angular momentum of the system in a linear way according to
\begin{equation}
\frac{J_{\rm init}-J_{\rm final}}{J_{\rm init}} = \gamma \frac{\Delta M}{M+ m},
\end{equation} 
where $J_{\rm init}$ resp. $J_{\rm final}$ is the angular momentum of the pre- and post-mass transfer binary respectively, and $m$ is the mass of the companion. We assumed $\gamma = 1.75$, see \citet{Nel01}.

    \begin{figure*}
    \centering
    \begin{tabular}{c c}
	\includegraphics[scale=0.4, angle=270]{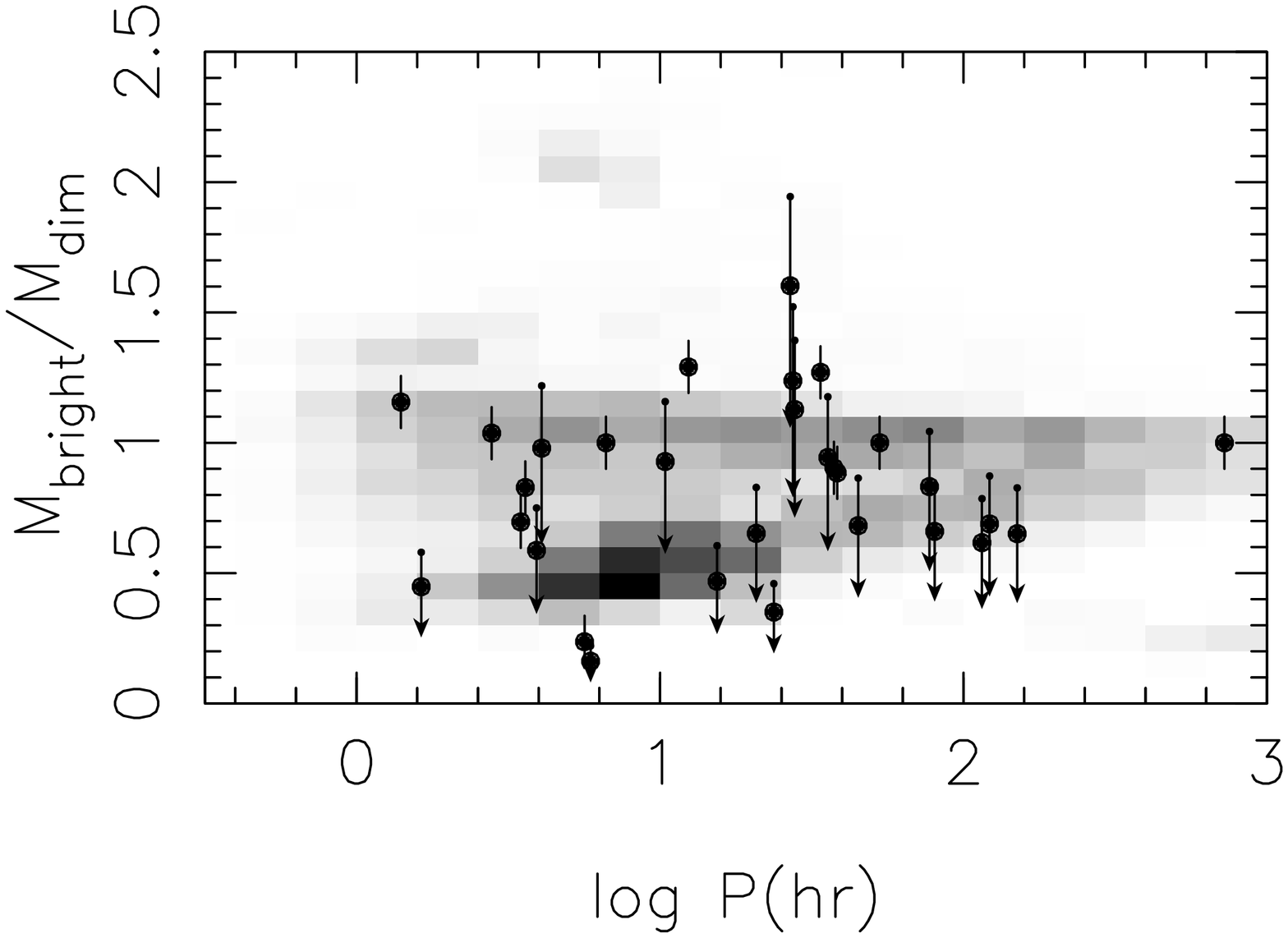} &
	\includegraphics[scale=0.4, angle=270]{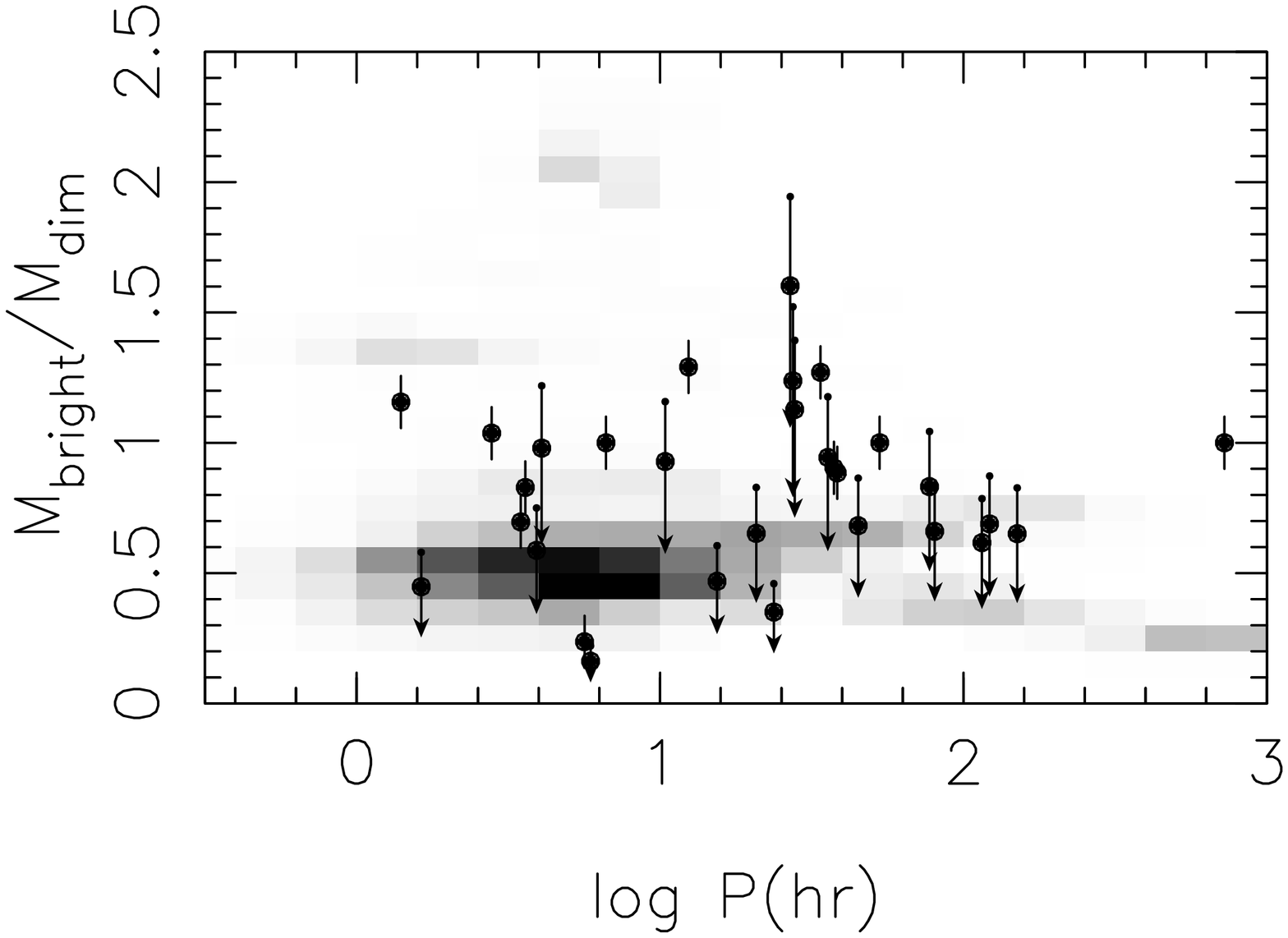} \\
	(a) & (b) \\
	\end{tabular}
    \caption{Simulated population of visible double white dwarfs  as a function of orbital period and mass ratio, where mass ratio is defined as the mass of the brighter white dwarf divided by that of the dimmer white dwarf. In the left model $\mga$ is used, in the right model $\maa$. The intensity of the grey scale corresponds to the density of objects on a linear scale. The same grey scale is used for both plots. Observed binary white dwarfs are overplotted with filled circles, see Fig. \ref{fig:pop_M1} for references.} 
    \label{fig:pop_q}
    \end{figure*}

We adopt two evolutionary models that differ in their treatment of the CE phase. In model $\alpha\alpha$ the $\alpha$-formalism is used to determine the outcome of every CE. For model $\mga$ the $\gamma$-prescription is applied unless the binary contains a compact object or the CE is triggered by a tidal instability (rather than dynamically unstable Roche lobe overflow, see Appendix \ref{appendix:stability}). Typically, the second CE (with a giant donor and white dwarf companion) is described by the $\alpha$-formalism, which gives consistent results when compared with the observations \citep{Nel00}. If the first phase of mass transfer is unstable, it typically evolves through a $\gamma$-CE. In model $\mga$ and $\maa$, if both stars are evolved when the CE develops, we assumed that both cores spiral-in \citep[see][]{Nel01}. The envelopes are expelled according to 

\begin{equation}
E_{\rm gr, \star don} + E_{\rm gr, \star comp} = \alpha (E_{\rm orb,init}-E_{\rm orb,final}),
\label{eq:dspi}
\end{equation} 

analogous to Eq.\ref{eq:alpha-ce}, where  $E_{\rm gr, \star don}$ represents the binding energy of the envelope of the donor star and $E_{\rm gr, \star comp}$ of the companion star.

The motivation for the alternative formalism is the large amount of angular momentum available in binaries with similar-mass objects. The physical mechanism behind the $\gamma$-description remains unclear however. Interesting to note here is that recently \citet{Woo10_Myk, Woo11} suggested a new evolutionary model to create DWDs. It differs from standard assumptions in the first phase of mass transfer. These authors find that mass transfer between a red giant and a main-sequence star can be stable and nonconservative. The effect on the orbit is a modest widening, with a result alike to the $\gamma$-description. 

\subsection{Impact on the population of double white dwarfs and type Ia supernova progenitors}
\label{sec:ce_obs}
Fig. \ref{fig:pop_q} shows the mass ratio of the visible population (see Sect. \ref{sec:seba_obs}) of DWDs versus the orbital period according to model $\mga$ and $\maa$. Overlayed with filled circles are the observed populations. 
For systems for which only a lower limit to the mass of the companion is known, we show a
plausible range of mass ratios of that system with an arrow. The arrow is drawn starting from the maximum mass ratio, which corresponds to an inclination of 90 degrees. It extends to a companion mass that corresponds to an inclination of 41 degrees. Within this range of inclinations there is a 75\% probability that the actual mass ratio lies along the arrow. The filled circles overplotted on the arrow indicate the mass ratio for the median for random orientations, i.e., 60 degrees. 

Using model $\maa$, the DWDs cluster around a mass ratio of $q\sim 0.5$, while model $\mga$ shows a wider range in mass ratio. This agrees better with the observed binaries. The different mass ratio distributions are inherent to the models and only slightly dependent on the CE efficiency. This is because in the first CE phase, the $\gamma$-CE allows for widening or very mild shrinkage of the orbit, whereas in the $\alpha$-prescription the orbit will always shrink. The resulting orbital separation determines when the secondary will fill its Roche lobe, and the corresponding core mass of the secondary, which determines the mass ratio distribution of the prospective DWD. 

\begin{figure*}
    \centering
    \begin{tabular}{c c}
	\includegraphics[scale=0.4, angle=270]{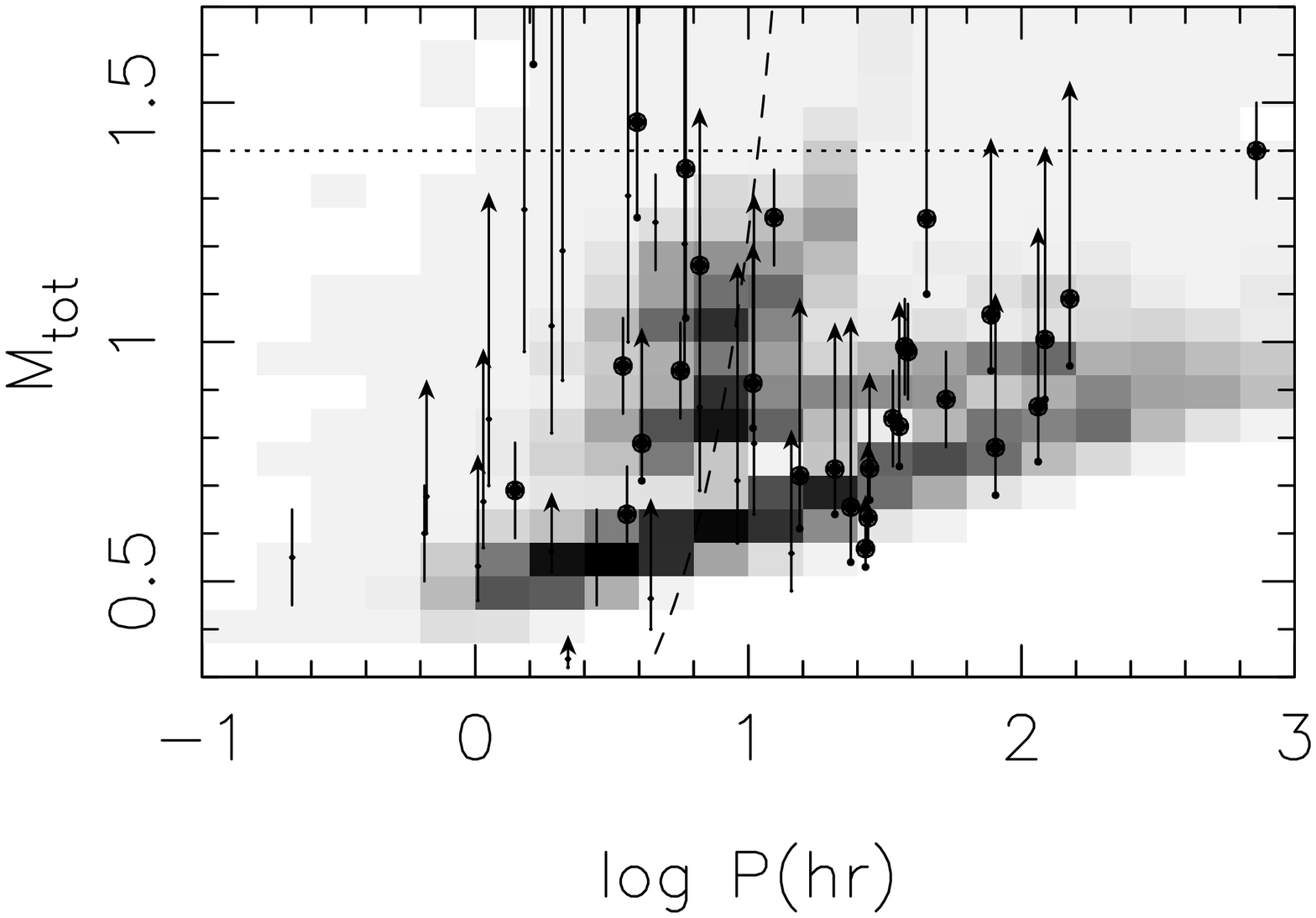} &
	\includegraphics[scale=0.4, angle=270]{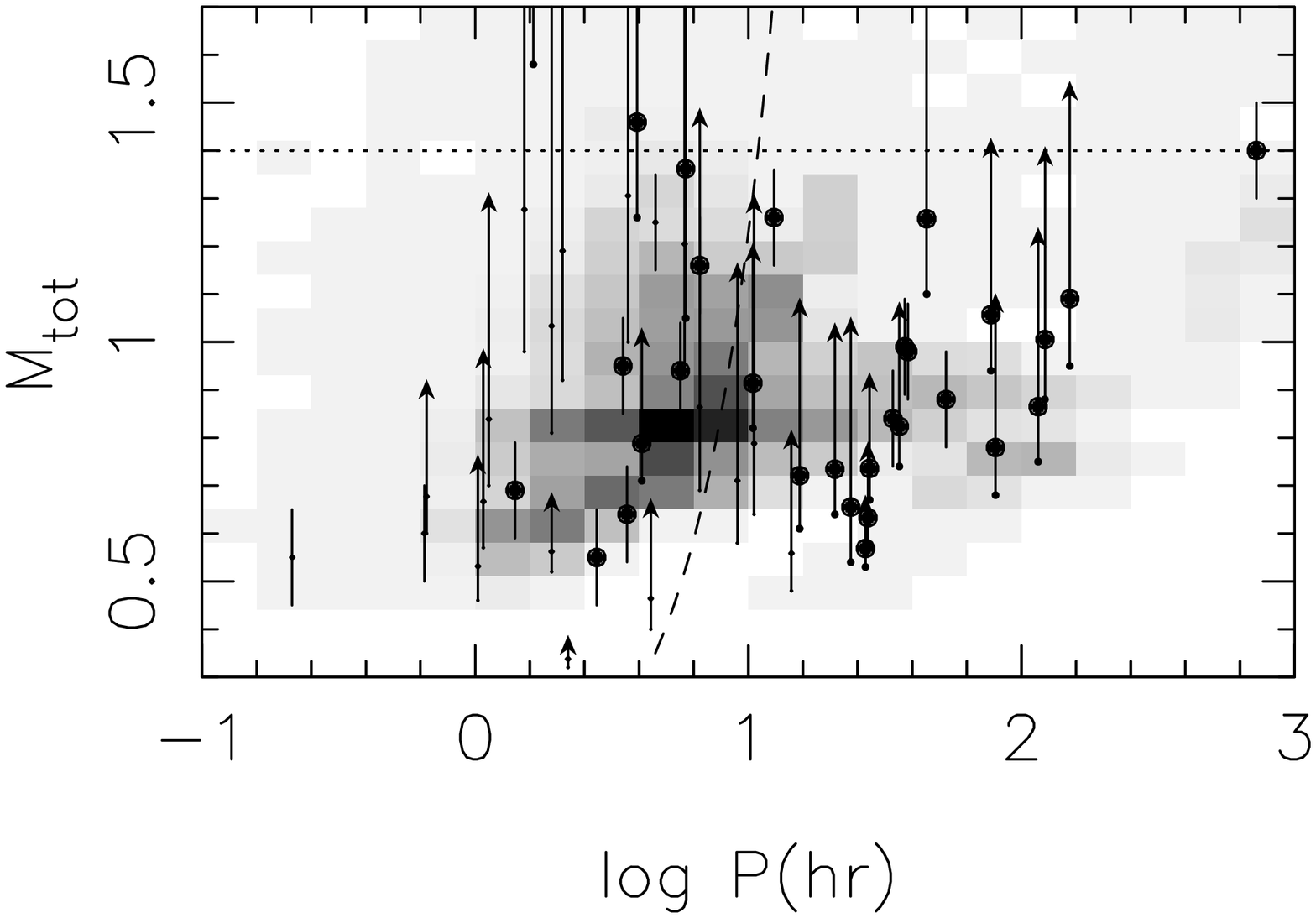} \\
	(a) & (b) \\
	\end{tabular}
    \caption{Simulated population of visible double white dwarfs as a function of orbital period and
    the combined mass of the two dwarfs. On the left the common envelope is parametrised according to model $\mga$, on the right according to model $\maa$ (see Sect. \ref{sec:ce}). The intensity of the grey scale corresponds to the density of objects on a linear scale. The same grey scale is used for both plots. Observed binary white dwarfs are overplotted with filled circles, see Fig. \ref{fig:pop_M1} for references. The Chandrasekhar mass limit is indicated with the dotted line. The dashed line roughly demarks the region in which systems merge within a Hubble time. Systems located to the left of the dashed line and above the dotted line are supernova type Ia progenitors in the standard picture. }
    \label{fig:pop_Mt}
    \end{figure*}

In Fig. \ref{fig:pop_Mt} the population of observed and simulated DWDs are shown as a function of combined mass of the two WDs for the two models of CE evolution. The left upper corner bounded by the dotted and dashed lines contains SNIa progenitors. In Fig. \ref{fig:pop_Mt} there are two systems that have a probability to fall in this region. These systems are the planetary nebulae nuclei with WD companions TS 01 (PN G135.9+55.9)\citep{Tov04, Nap05, Tov10} and V458 Vul \citep{Rod10}. An immediate precursor of a DWD that is possibly a progenitor candidate for a SN Ia via the DD channel has also been observed; a subdwarf with a white dwarf companion, KPD 1930+2752 \citep{Max00, Gei07}. 

In our model of the visible population of DWDs (see Sect. \ref{sec:seba_obs}), the percentage of merging DWDs with a total mass exceeding the Chandrasekhar is 1.2\% for model $\mga$ and 4.3\% for model $\maa$. Including only double CO WDs, the percentage is 0.9 and 2.9\%, respectively. Because the number of observed close DWDs until today is low, we do not expect to observe many SNIa progenitors \citep[see also][]{Nel01}. 
Therefore a comparison of the SNIa progenitors with population synthesis by a statistical approach is unfortunately not yet possible. We find it important to compare the observed close DWD population with the simulated one, since these systems go through similar evolutions and are strongly influenced by the same processes. Although the observed population mostly consists of He DWDs and He - CO DWDs instead of CO DWDs required for SNIa progenitors, at this time the population of all close DWDs are the closest related systems that are visible in bulk.

\section{Evolutionary paths to supernova type Ia from the double degenerate channel} 
\label{sec:path} 
In this section we discuss the most common binary scenarios that leads to a potential supernova type Ia in the DD channel. We assume that every merger of two carbon/oxygen white dwarf with a mass exceeding 1.4\Msolar\ will lead to a supernova. The contribution of merging systems that contains a helium white dwarf that surpasses the Chandrasekhar mass is negligible. In the canonical scenario a DWD is formed through two consecutive CEs. This we label the 'common envelope channel'. In accordance with \citet{Men10}, we find that there are other channels that can lead to a SNIa as well. We find that the common envelope scenario can account for less than half of the supernova progenitors in a single burst of star formation, 34\% for model $\mga$ and 45\% for model $\maa$. We distinguish between three scenarios labelled 'common envelope', 'stable mass transfer' and 'formation reversal'. The names of the first two tracks refer to the first phase of mass transfer, whereas 'formation reversal' applies to the reversed order in which the two white dwarf are formed, see Sect. \ref{sec:formation_reversal}). 
The stable mass transfer channel accounts for 52\%  and 32\% assuming model $\mga$ and $\maa$, respectively, for a single burst of star formation. The formation reversal channel accounts for a lower percentage of all SNIa, 14\% for model $\mga$ and 23\% for model $\maa$ for a single starburst. 
Note that the importance of the stable mass transfer channel strongly depends on the assumed amount of mass loss and angular momentum loss.

In population synthesis studies all known information about binary evolution is combined, and different evolutionary paths emerge out of these quite naturally. As noted by \citet{Men10}, the significant contribution to the SNIa rate from other channels than the common envelope channel complicates the use of analytical formalisms for determining the distribution of SNIa delay times. The SNIa delay time of a binary is the time of the SNIa since the formation of the system. This is commonly used to compare observational and synthetic rates to constrain different physical scenarios 
\citep[e.g.][see also Sect. \ref{sec:dtd} in this paper]{Yun00, Rui09, Men10}. During a CE phase the companion is assumed to be hardly affected e.g. by accretion. If this is not the case, as in stable mass transfer, the assumption that the formation timescale of the DWD is approximately the main-sequence lifetime of the least massive component is not valid any more. Furthermore, the in-spiral timescale from DWD formation to merger due to gravitational waves is strongly dependent on the orbital separation at DWD formation. This can be very different for systems that undergo stable mass transfer instead of a CE evolution. Concluding, the delay time, which is the sum of the DWD formation and in-spiral timescale can be significantly different when these tracks are not properly taken into account.

\begin{figure}
\centering
\includegraphics[scale=0.425, angle = 270]{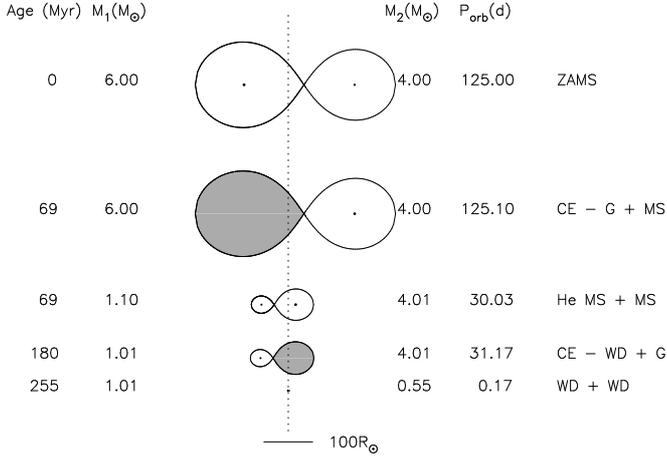}
\caption{Evolutionary track for the merger of two carbon/oxygen white dwarfs of a combined mass that exceeds the Chandrasekhar mass. In this scenario the first phase of mass transfer is dynamically unstable which results in a common envelope. In this figure we show a representative example of model $\mga$, see Sect. \ref{sec:ce}. ZAMS stands for zero age main sequence and CE for common envelope. G is a giant star, MS a main-sequence star, He MS a helium MS and WD a white dwarf.}
\label{fig:track_dyn}
\end{figure}

\subsection{Common envelope channel}
In the canonical path, both stars lose their hydrogen envelopes through two consecutive common envelopes. An example of a typical evolution is shown in Fig. \ref{fig:track_dyn}. In this example two zero-age main-sequence stars of 6\Msolar\ and 4\Msolar\ are in an orbit of 125 days. When the initially more massive star (hereafter primary) ascends the giant branch, it fills its Roche lobe and a CE commences. 
The primary loses its hydrogen envelope, but does not become a WD immediately and a helium star is born. 
%In this example the primary fills its Roche lobe again as a helium giant, leading to a small decrease of the mass of the primary and small increase in the orbital separation. 
The primary becomes a white dwarf of about solar mass. 
When the initially less massive star (hereafter secondary) evolves off the main sequence and its radius significantly increases, another common envelope occurs. As a result, the orbit shrinks. The secondary evolves further as a helium star without a hydrogen envelope until it eventually turns into a white dwarf. 
For model $\maa$ the orbit decreases more severely in the first phase of mass transfer. Therefore the initial periods in this channel are higher, by a factor $\sim 1.5-3$ and the primaries are typically more evolved giants when the donors fill their Roche lobes.  
In this evolution channel both the primary and secondary can fill their Roche lobes as helium giants.  
If this happens, mass transfer is usually dynamically stable, but the effect on the orbit is small.

A variation of this evolution can occur when the secondary has reached the giant stages of its evolution when the primary fills its Roche lobe. This happens for systems of nearly equal masses. We assume both stars lose their envelope in the CE phase according to Eq. \ref{eq:dspi}, in which the orbit is severely decreased. This variation contributes 23\% of the systems in the common envelope channel for model $\mga$ and 10\% for model $\maa$.    

With the $\alpha$-CE prescription it is likely to have another variation on the evolution, in which the primary becomes a white dwarf immediately after the first phase of mass transfer. This can happen when the primary fills its Roche lobe very late on the asymptotic giant branch when the star experiences thermal pulses. These systems have initial periods that are a factor 5 lger than in the standard CE channel using model $\maa$. 
This subchannel contributes 43\% to the CE channel for model $\maa$. When using the $\mga$-model for the CE, the contribution from this subchannel is 20\%. However, these systems are not formed through a standard $\gamma$-CE because the orbit does not shrink severely enough to obtain a significant contribution. Instead these systems are formed through a double-CE as described by Eq.\ref{eq:dspi}. For model $\maa$ the double-CE mechanism is important in only 18\% of the 43\%.

\subsection{Stable mass transfer channel}
\begin{figure*}
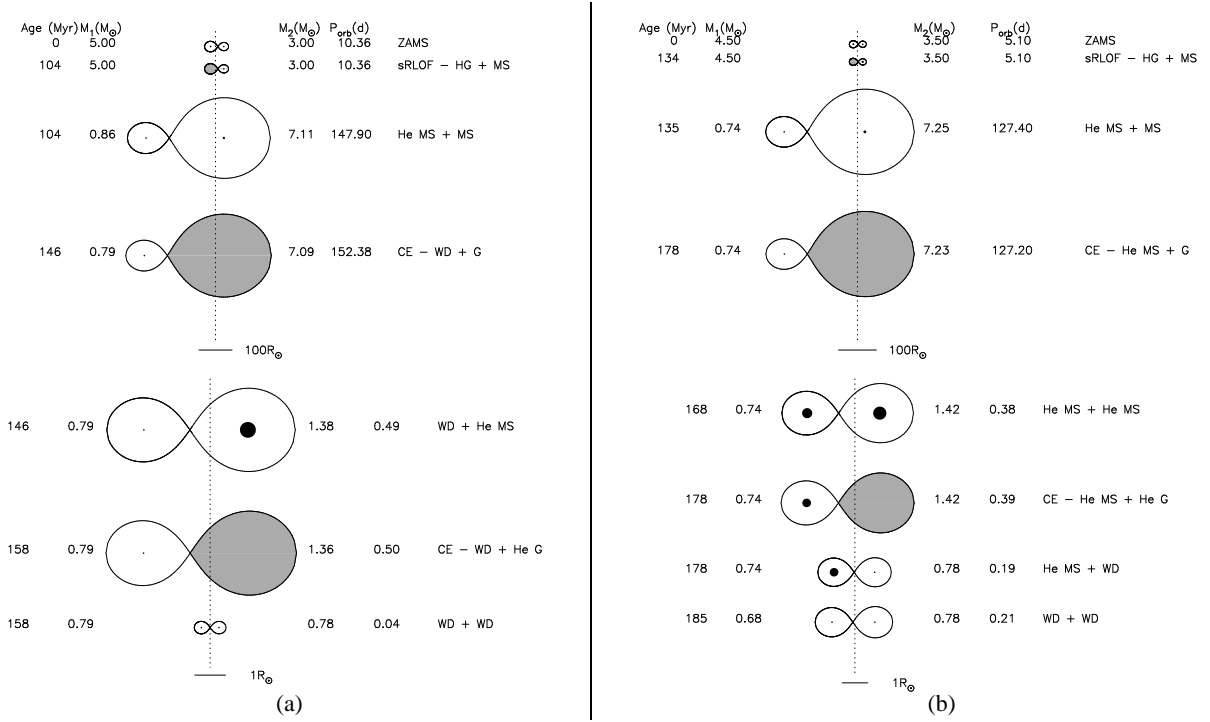

\centering
\begin{tabular}{cc|cc}
\includegraphics[scale=0.325, angle=270]{rochelobes_ther1.eps} &&&
\includegraphics[scale=0.325, angle=270]{rochelobes_rev1.eps} \\
%\quad \quad CE & && \quad \quad CE \\
\includegraphics[clip=true, trim= 10mm 0mm 0mm 0mm, scale=0.325, angle=270]{rochelobes_ther2.eps} &&&
\includegraphics[clip=true, trim= 10mm 0mm 0mm 0mm, scale=0.325, angle=270]{rochelobes_rev2.eps} \\
\quad \quad (a) &&& \quad \quad (b) \\
\end{tabular}
\caption{Two evolutionary tracks for the merger of two carbon/oxygen white dwarfs of a combined mass exceeding the Chandrasekhar mass. In these scenarios the first phase of mass transfer is dynamically stable. On the left an example of the stable mass transfer channel is shown, on the right the formation reversal channel. In the latter scenario the first phase of mass transfer is dynamically stable which results in a low-mass helium-star with a long lifetime. The initially less massive star becomes the first formed white dwarf. The top and bottom parts of the figure have different scales due to a common envelope phase, denoted as CE in the figure. Abbreviations are as in Fig. \ref{fig:track_dyn}. Additionally sRLOF stands for stable Roche lobe overflow, HG is a Hertzsprung-gap star and He G a helium giant. }
\label{fig:track_ther}
\end{figure*}

In this channel the initial masses of the stars and the initial orbits are smaller than for the common envelope channel. Typical values are a primary mass of 5\Msolar, a secondary mass of 3\Msolar\ and an orbital separation of 40\Rsolar\ (assuming a circular orbit). The primary fills the Roche lobe as a Hertzsprung gap star 
and mass transfer occurs stably. Which fraction of transferred mass is actually accreted by the secondary and how much is lost from the system depends on the mass and radius of the secondary and the secondary's Roche lobe (see Appendix \ref{appendix:accretion} for more details). In Fig. \ref{fig:track_ther}a an example of a typical evolution is shown. 
When the secondary fills its Roche lobe, a CE commences. In this channel the tidal instability (see Appendix \ref{appendix:stability}) is important. In one third of the systems the CE occurs because of a tidal instability, the other part is caused by a dynamical instability. The secondary turns into a hydrogen-deficient helium-burning star in a system in which the period has decreased by one or two orders of magnitude. 
As in the previous channel, the primary and secondary can fill their Roche lobe as helium giants. 
If the primary fills its Roche lobe, mass transfer is usually dynamically stable and has little effect on the orbit. If the secondary fills its Roche lobe, mass transfer can be stable or unstable.
In the example of Fig. \ref{fig:track_ther}a when the secondary fills its Roche lobe again as it ascends the helium giant branch, the mass transfer is unstable and the orbital separation decreases by a factor $\sim 5$.

\subsection{Formation reversal channel}
\label{sec:formation_reversal}
We present a scenario in which in the first mass transfer a helium star (sdB star) is formed that becomes a white dwarf only after the companion has become a white dwarf. 
\footnote{This track is a close analogy of the track proposed by \citet{Sip04} regarding recycled pulsars. In the scenario proposed by these authors the end states of the two components are reversed, resulting in a neutron star that forms prior to a black hole. However, in our scenario the name 'formation reversal' applies to the evolutionary timescales of the primary and secondary. Although the primary first evolves off the main sequence, the secondary becomes a remnant first.} 
A typical example of an evolution like this is shown in Fig. \ref{fig:track_ther}b. The first phase of mass transfer is stable, like the stable mass transfer track. However, the resulting helium stars in this channel have low masses in the range of 0.5-0.8\Msolar\ and long lifetimes of $\sim 10^8$ yr. 
The first mass transfer occurs approximately conservatively. As a consequence, the subsequent evolution of the high-mass secondary (5-8\Msolar) accelerates. When the secondary fills its Roche lobe, mass transfer is tidally unstable (see Appendix \ref{appendix:stability}). The secondary loses its hydrogen and helium envelope in two consecutive CEs and becomes a white dwarf. Subsequently, the original primary evolves of the helium main-sequence and becomes a white dwarf.

To our knowledge, this track has not been studied in detail before. Therefore we evaluated this track by performing detailed numerical calculations using the \texttt{ev} binary stellar-evolution code originally developed by Eggleton \citep[][and references therein]{Egg71, Egg72, Yak05} and updated as described in \citet{Pol95} and \citet{Gle08}. The code solves the equations of stellar structure and evolution for the two components of a binary simultaneously. The simulation showed that indeed the evolution of the secondary can be accelerated through accretion so that the secondary can stop helium burning prior to the primary. 

\begin{figure}
\includegraphics[width=\columnwidth]{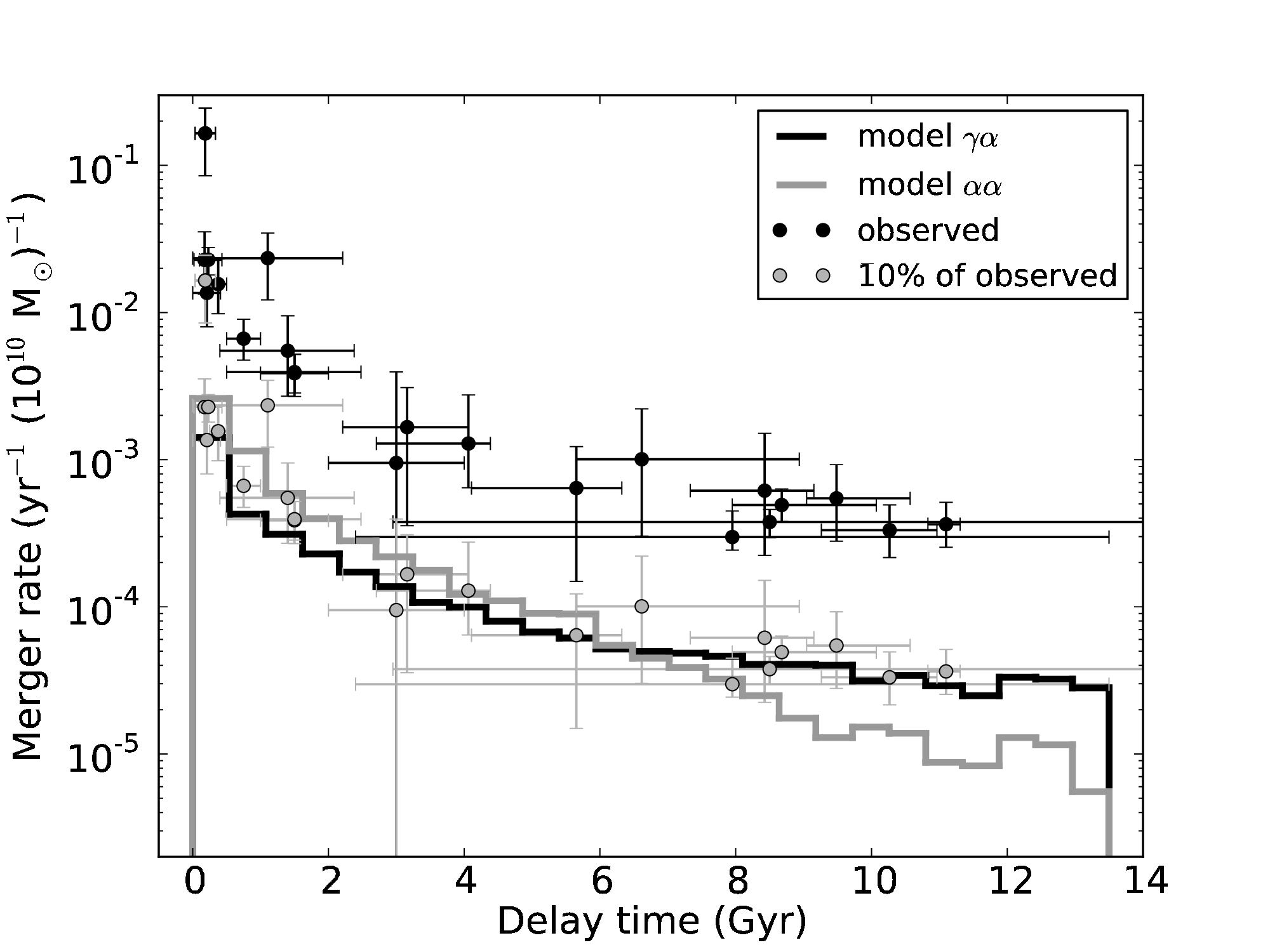}
\caption{Merger rate of double carbon/oxygen white dwarfs with a total mass above the Chandrasekhar mass as a function of delay time. Rates are in yr$^{-1}$ per 10$^{10}$\Msolar\ formed stellar mass of the parent galaxy. Solar metallicity ($Z=0.02$) is assumed. Delay times are shown for two different prescriptions of the CE phase. In black we plot model $\gamma\alpha$ and in grey model $\alpha\alpha$, see Sect. \ref{sec:ce}. Overplotted with black circles are the observed values of the SNIa rate of \citet{Tot08}, \citet{Mao10}, \citet{Mao10b} and \citet{Mao11} \citep[see][for a review]{Mao11b}. For comparison the grey circles show the observations scaled down by a factor 10.}
\label{fig:dtd_alpha}
\end{figure}

\begin{figure}
\includegraphics[width=\columnwidth]{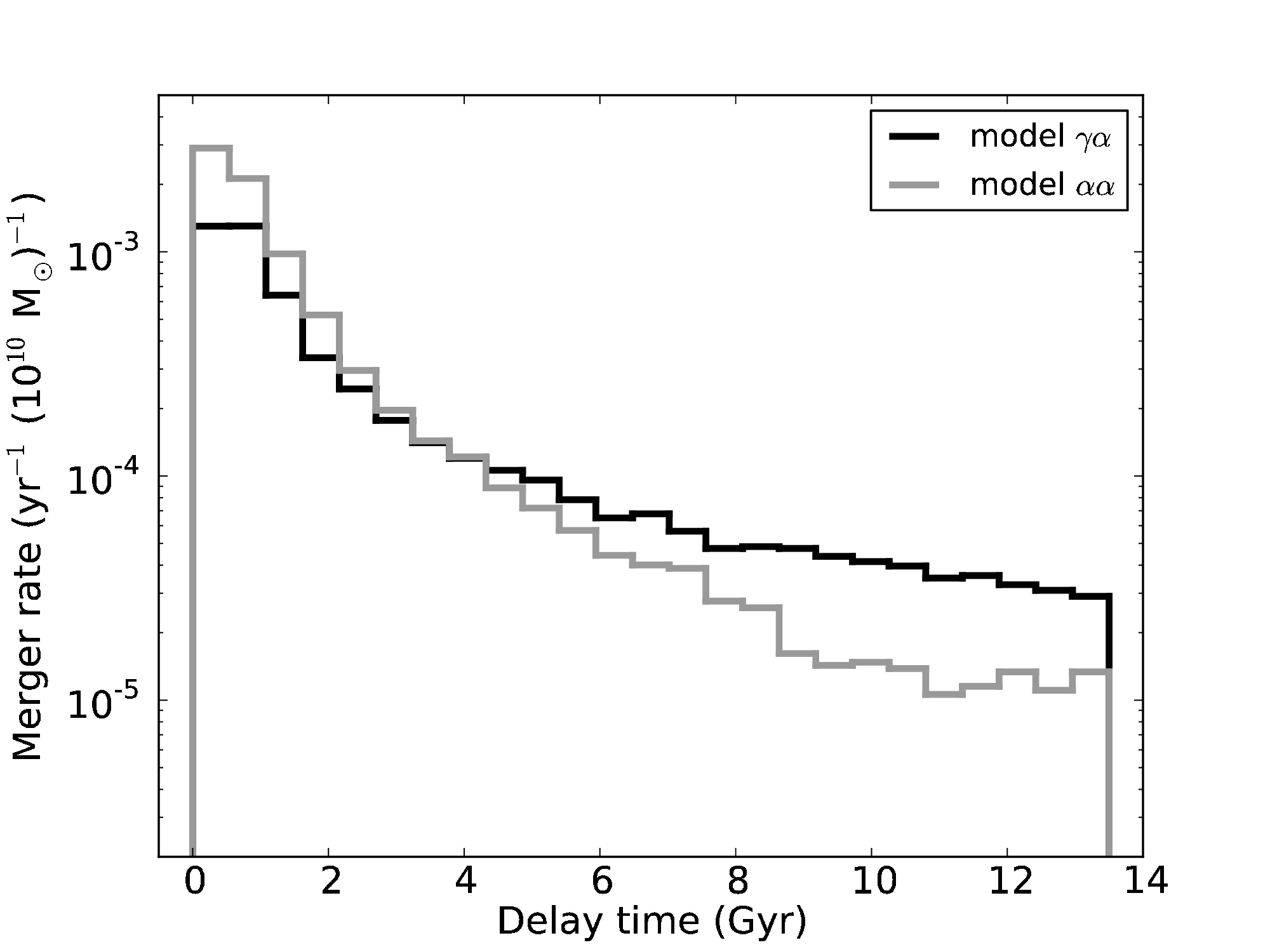}
\caption{Merger rate of double carbon/oxygen white dwarfs with a total mass above the Chandrasekhar mass as a function of delay time for a metallicity of 0.001. Rates are in yr$^{-1}$ per 10$^{10}$\Msolar\ per formed stellar mass of the parent galaxy. 
Delay times are shown for two different prescriptions of the CE phase. In black model we plot $\gamma\alpha$ and in grey model $\alpha\alpha$, see Sect. \ref{sec:ce}.}
\label{fig:dtd_z}
\end{figure}

\section{Delay time distribution}
\label{sec:dtd}
One way to constrain the population of SNIa progenitors is through the delay time distribution (DTD), where the delay times is the time between the formation of the binary system and the SNIa event. In a simulation of a single burst of star formation the DTD gives the SNIa rate as a function of time after the starburst. The DTD is linked to the nuclear timescales of the progenitors and the binary evolution timescales up to the merger. We assumed a 50\% binary fraction and initial parameters are distributed according to the distributions described in Table \ref{tbl:init_param}.

In Fig. \ref{fig:dtd_alpha} we compare the delay time distribution for the two different models of CE evolution. The sharp cut-off near 13.5 Gyr is artificial, because evolution was only allowed to proceed for 13.5 Gyr. The delay time distribution shows that these mergers are expected to take place in young as well as old populations. The peak in the supernova Ia rate is at $\sim$150 Myr for both models. The  median delay time is 0.7 Gyr for model $\maa$ and 1.0 Gyr for model $\mga$. 
The normalisations of the delay time distribution of model $\maa$ and $\mga$ are comparable. The time-integrated number of SNe Ia per unit formed stellar mass is $2.0\cdot 10^{-4} \Mo^{-1}$ and $3.3 \cdot 10^{-4} \Mo^{-1}$ for model $\gamma\alpha$ and $\alpha\alpha$, respectively. From the Lick Observatory Supernova Search, \citet{Mao11} inferred a value of $2.3\pm 0.6 \cdot 10^{-3}\ \Mo^{-1}$ ,which is a factor 7-12 higher than the predictions from our models.  

The morphologies of the DTD of model $\mga$ and $\maa$ resemble each other in that they show a strong decline with delay time, although with a slightly different slope. The $\maa$ model shows higher rates at short delay times, whereas the rate for model $\mga$ shows higher rates at long delay times. This is because the $\alpha$-CE causes a stronger decrease of the orbital separation than the $\gamma$-CE in the first phase of mass transfer. The observed rates from \citet{Tot08}, \citet{Mao10}, \citet{Mao10b}, and \citet{Mao11} \citep[see][for a review]{Mao11b}, shown in Fig. \ref{fig:dtd_alpha}, are much higher than the predicted rates from both models. To compare the morphological shapes of the DTDs more easily, we scaled down the observations by a factor 10 in Fig. \ref{fig:dtd_alpha} in light grey. The shape of the observed DTD fits the synthetic DTDs well. At long delay times $>6$ Gyr, the flattening of the DTD is better reproduced by the $\mga$-model.

We have a last remark about Fig. \ref{fig:dtd_alpha}, about the datapoint from \citet{Mao10} at 185Myr and a rate of 0.165$\peryr\ (10^{10} \Mo)^{-1}$. If this datapoint is true, it could indicate a steep rise of the delay time distribution at the shortest delay times. Neither model $\mga$, or $\maa$ reproduces the steep rise indicated by this point. At short delay times the contribution to the SNIa rate from other channels might be significant, for instance the contribution from helium donors in the SD channel. \citet{Wan09}, \citet{Rui09} and \citet{Cla11} showed that at delay times of $\sim 100$Myr, the DTD from helium donors in the SD channel peaks, although rates at this delay time vary between $10^{-4}-10^{-2} \peryr\ (10^{10} \Mo)^{-1}$. Hydrogen donors in the SD channel are a possible contributor to the SNIa rate as well, but there is a strong disagreement over the DTD from this channel, \citep[see for example][for an overview]{Nel12}. In that paper it was shown that the simulated peaks of the DTDs lie anywhere between 0.1 to 3 Gyr and the peak rates vary between $10^{-6}-10^{-2} \peryr\ (10^{10} \Mo)^{-1}$.

If we do not assume an instantaneous burst of star formation, but instead convolve the DTD with a star formation rate, we can estimate the SNIa rate from double degenerates for spiral galaxies like the Milky Way. If we assume a Galactic star formation rate as in \citet{Nel04} based on \citet{Boi99}, model $\maa$ gives $8.3\cdot 10^{-4}$ SNIa yr$^{-1}$. Model $\gamma\alpha$ gives a Galactic rate of $5.8\cdot 10^{-4}$ SNIa yr$^{-1}$. The reason for the relatively high Galactic rate for model $\mga$ in comparison with model $\maa$ relative to the integrated rates is that the peak of star formation occurs at long delay times where the DTD of model $\mga$ dominates over model $\maa$. The empirical SN Ia rate from Sbc-type galaxies like our own \citep{Cap01} is $4 \pm 2 \cdot 10^{-3} \peryr$, which is a factor $\sim 5-7$ higher than the simulated rates.  

When convolving the DTD with a star formation history, one should also take into account a metallicity dependence of the stars. To study the effect of metallicity on the SNIa rate, we simulated a delay time distribution from a single burst of stars of metallicity $Z=0.001$. The important part of the DTDs in this respect are the long delay times because this is where the fraction of metal poor stars is highest.  
The DTD of model $\mga$ for $Z=0.001$ is lower at long delay times and higher at short delay times than the same CE model for solar metallicities.
The time-integrated number of SNe Ia per unit formed stellar mass is $2.8\cdot 10^{-4} \Mo^{-1}$ for model $\mga$. This is an increase of $\sim$60\% with respect to $Z=0.02$. 
The integrated rate for model $\maa$ for $Z=0.001$ is $4.2\cdot 10^{-4} \Mo^{-1}$, which is an increase of $\sim$30\% with respect to $Z=0.02$. The DTD for $Z=0.001$ is roughly similar in morphological shape to that for $Z=0.02$, see Fig. \ref{fig:dtd_z}.
The DTDs of both metallicities for model $\maa$ are similar at long delay times, and consequently the effect on the Galactic SNIa rate is expected to be marginal. 

\section{Population of merging double white dwarfs}
\label{sec:pop}

\begin{figure*}
    \centering
    \begin{tabular}{c c}
	\includegraphics[scale=0.4]{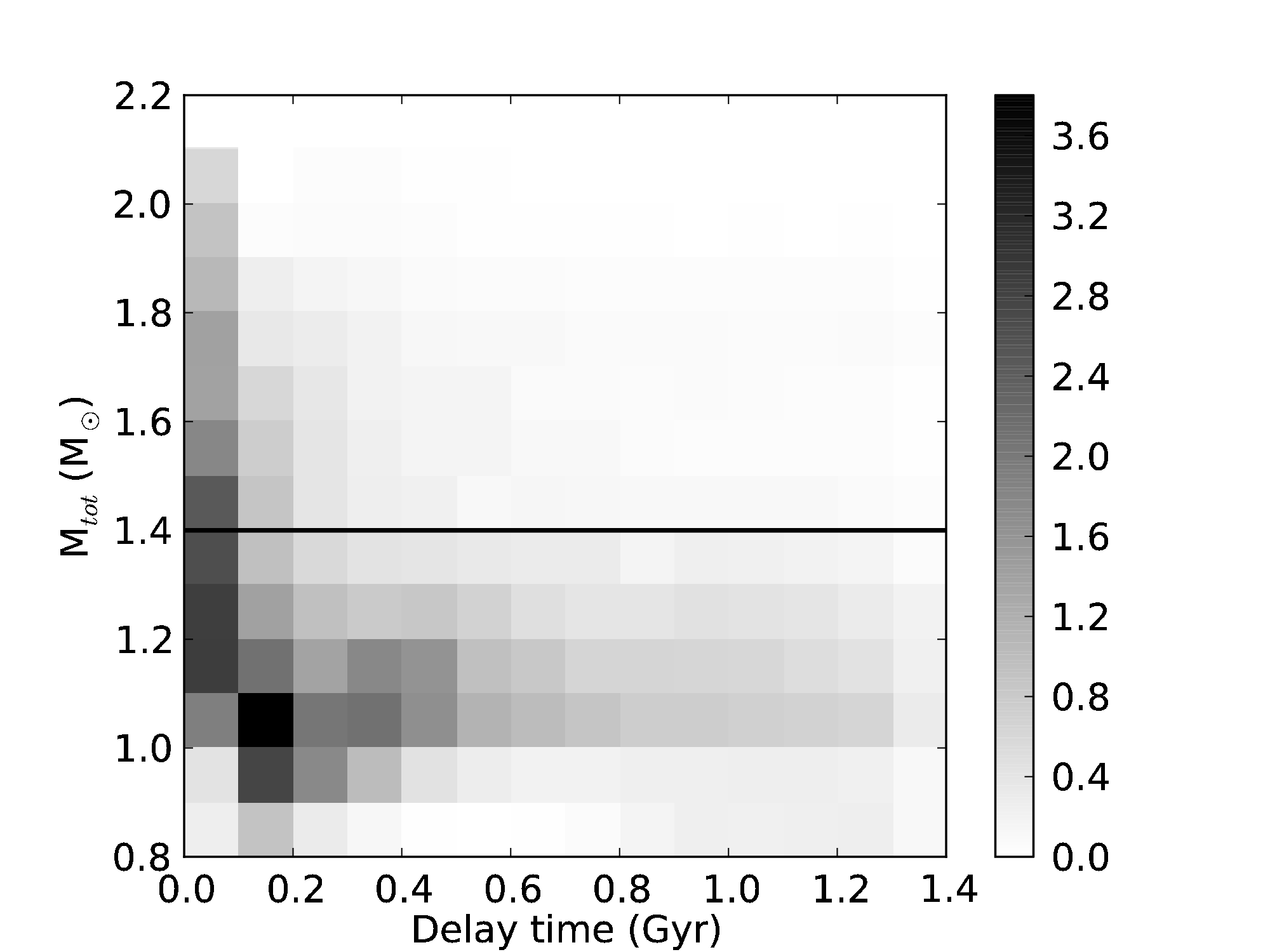} &
	\includegraphics[scale=0.4]{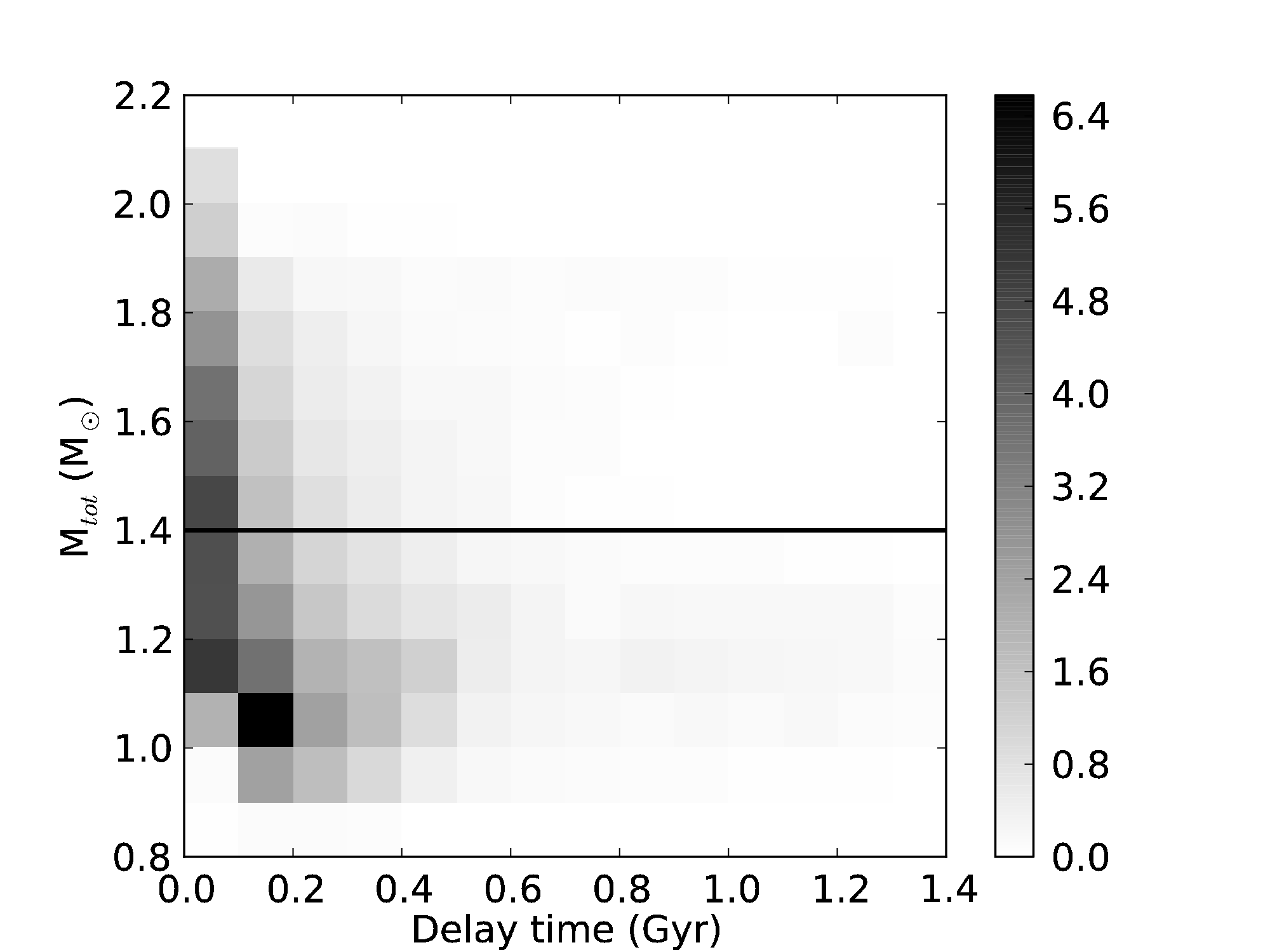} \\
	(a) & (b) \\
	\end{tabular}
    \caption{Simulated distribution of the population of merging double CO white dwarfs from a single burst of star formation as a function of delay time and total mass of the system. On the left model $\mga$ is used for the common envelope parametrisation, on the right model $\maa$ (see Sect. \ref{sec:ce}). The intensity of the grey scale corresponds to the density of objects on a linear scale in units of number of systems per $10^5\Mo$. The black line corresponds to a combined mass of 1.4\Msolar.}
        \label{fig:pop_gamma_tdelay}
    \end{figure*}

In this section we discuss the properties of the population of SNIa progenitors from merging DWD and place it in the context of recent studies of the SNIa explosion itself. Fig. \ref{fig:pop_gamma_tdelay} shows the combined mass of the system as a function of delay time for merging CO DWDs. It shows that for classical SNIa progenitors, the number of merging events decreases with time and that the number decreases faster with time for model $\maa$ than for model $\mga$, as discussed in Sect. \ref{sec:dtd}. Moreover, the figure shows that mergers near the Chandrasekhar mass are most common, independent of delay time. 
 
\citet{Fry10} showed that if super-Chandrasekhar mergers of CO DWDs of $\sim 2\Mo$ produce thermonuclear explosions, the light curves are broader than the observed SNIa sample. These authors argued that these mergers cannot dominate the current SNIa sample. We find indeed in both models that mergers with combined masses $\sim 2\Mo$ are much less common than mergers in systems with a combined mass near the Chandrasekhar mass limit. 

\begin{figure*}
    \centering
    \begin{tabular}{c c}
	\includegraphics[scale=0.4]{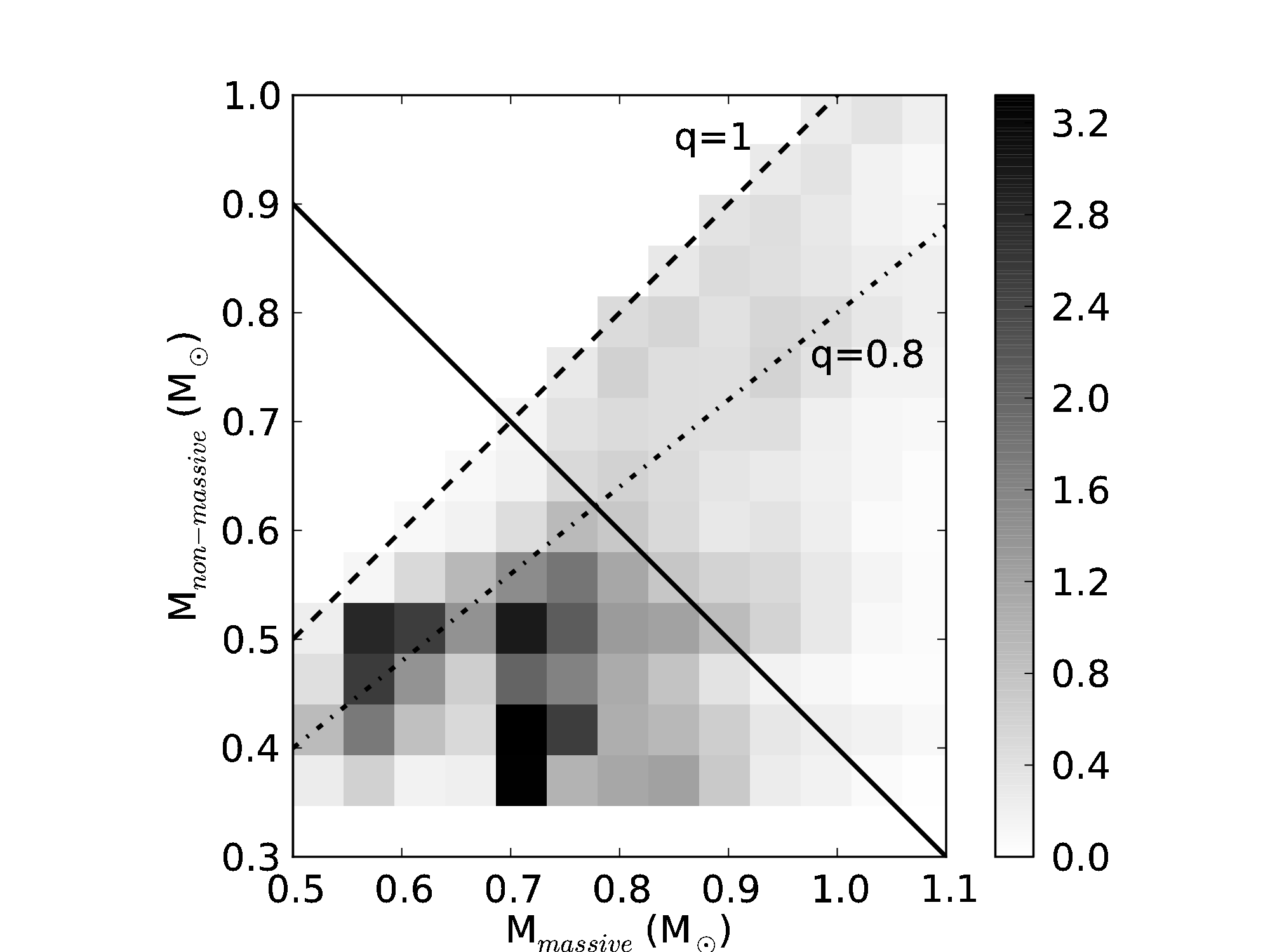} &
	\includegraphics[scale=0.4]{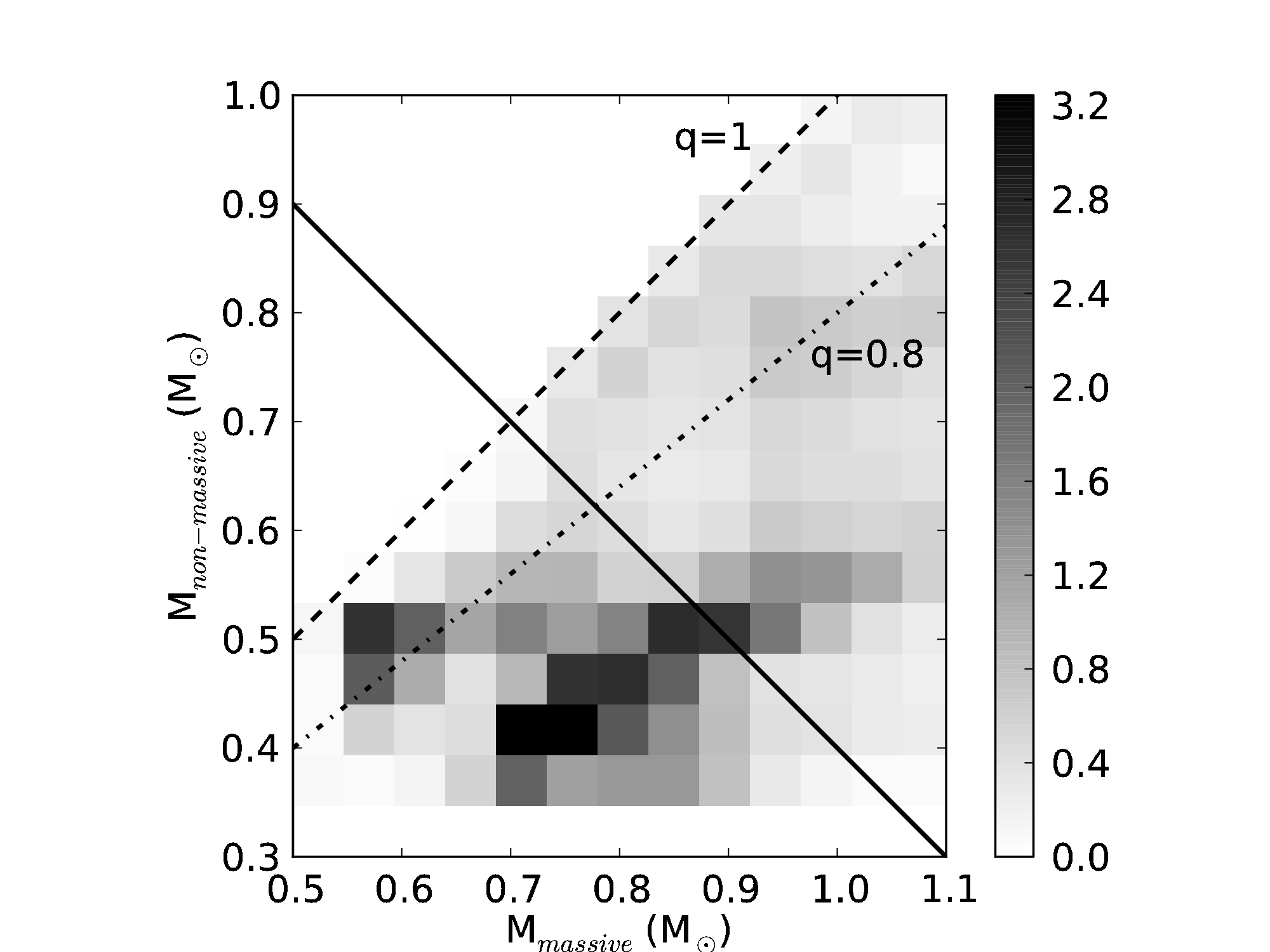} \\
	(a) & (b) \\
	\end{tabular}
    \caption{Simulated population of merging double CO white dwarfs from a single burst of star formation as a function of the masses of the two white dwarfs. $M_{\rm massive}$ is the mass of the most massive white dwarf, $M_{\rm non-massive}$ corresponds to the least massive white dwarf. On the left model $\mga$ is used for the common envelope parametrisation, on the right model $\maa$ (see Sect. \ref{sec:ce}). The intensity of the grey scale corresponds to the density of objects on a linear scale in units of number of systems per $10^5\Mo$. To increase the contrast, we placed an upper limit on the intensity, which effects only one bin for model $\mga$ and two bins for model $\maa$. The black solid line corresponds to a combined mass of 1.4\Msolar. The dashed and dashed-dotted line correspond to a mass ratio $q=m/M$ of 1 and 0.8, respectively. The mass ratio distribution, which is important e.g. in the scenarios proposed by \citet{Pak10} and \citet{Ker10} are very different for model $\mga$ and $\maa$.}
    \label{fig:pop_gamma_mass}
    \end{figure*}

Where \citet{Fry10} studied a merger of a $1.2\Mo$ CO WD with a $0.9\Mo$ CO WD, \citet{Pak10} focused on mergers of nearly equal mass WDs. In their scenario both WDs are distorted in the merger process and the internal structure of the merger remnant is quite different. \citet{Pak10} argued that these mergers become hot enough to ignite carbon burning if the WD masses exceed $M \gtrsim 0.9M_{\odot}$. They found that these systems resemble subluminous SNIa such as SN 1991bg. \citet{Li01} found 1991bg-like supernovae account for $16\pm6 \%$ of all SNIa. From an improved sample \citet{Li11} found a percentage of $15.2^{6.8}_{5.9}$. If we assume that 1991bg-like events account for 15\% of all SNIa and the time-integrated of all SNIa types is $2.3 \pm 0.6 \cdot 10^{-3} \Mo^{-1}$ \citep{Mao11}, the time-integrated rate should be $3.5 \pm 0.9\cdot 10^{-4}\ \Mo^{-1}$. When we include the error of 6\% on the fraction of 1991bg-like events with respect to all SNIa, the rate is $3.5 \pm 1.6 \cdot 10^{-4} \Mo^{-1}$. 
 If we assume in a simplistic way that all CO DWD mergers of $q=M_2/M_1 > 0.92$ and $M_1>0.9$ (where $M_1$ is the most massive WD and $M_2$ the least massive WD) would lead to a 1991bg-like event, the time-integrated number of events is $2.3 \cdot 10^{-5} \Mo^{-1}$ according to model $\mga$ and $1.8\cdot 10^{-5} \Mo^{-1}$ assuming model $\maa$. While the SNIa rate from the classical progenitors from model $\mga$ is comparable to that of model $\maa$, the population of DWDs is very different. In Sect. \ref{sec:ce_obs} we showed that the type of CE parametrisation introduces a bias in the mass ratio distribution of observed DWDs, which mostly consist of (double) He DWDs and He-CO DWDs. In Fig. \ref{fig:pop_gamma_mass} we show that this is also the case for the population of merging CO DWDs. Although the mass ratio distribution is not important for the standard DD scenario, it is important for the scenario proposed by \citet{Pak10}. If the standard scenario and the scenario proposed by \citet{Pak10} hold, SN 1991bg-like events are more common in model $\mga$. We have to make a side remark on the expected delay times of this scenario. The median delay times are 180 Myr for model $\mga$ and 150 Myr for model $\maa$. The timescales are short because generally, more massive WDs have more massive progenitor stars, whose evolutionary timescales are short compared to those of less massive stars. Observations show, however, that subluminous SNIa are associated with old stellar populations $\sim 5-12$Gyr \citep{How01}.

In our simulations the majority of merging CO DWDs have combined masses below the Chandrasekhar mass, see Fig. \ref{fig:pop_gamma_tdelay}. Sub-Chandrasekhar models have long been proposed in order to raise total number of SNIa to match observations. \citet{Sim10} found that if sub-Chandrasekhar WDs can be detonated, especially in the range  $\sim 1.0-1.2 \Mo$, the explosions match several observed properties of SNIa reasonably well. 
In the hypothetical situation that all double CO WDs that merge lead to an SNIa event, the integrated rate is $8.3 \cdot 10^{-4} \Mo^{-1}$ for model $\mga$ and $9.3\cdot 10^{-4} \Mo^{-1}$ for model $\maa$. This is still a factor 3 lower than the observed rate of $2.3 \pm 0.6 \cdot 10^{-3} \Mo^{-1}$ \citep{Mao11}. Only if we assume that all mergers between a CO WD and a CO or He WD lead to an SNIa, the rates of $1.6 \cdot 10^{-3} \Mo^{-1}$ and $2.1 \cdot 10^{-3} \Mo^{-1}$ for model $\mga$ and $\maa$, respectively, match the observed rate.

The challenge for sub-Chandrasekhar models is how to detonate the white dwarf. A scenario for this was recently suggested by \citet{Ker10}. In this scenario two CO WDs with nearly equal masses merge. The merger remnant itself is too cold and insufficiently dense to produce an SNIa by itself, as noted by \citet{Pak10}. 
\citet{Ker10} proposed that accretion of the thick disk that surrounds the remnant leads to an SNIa through compressional heating. If we simplistically assume that every merger of a double CO DWD with $q>0.8$ and $M_1+M_2 < 1.4$ leads to an SNIa, the time-integrated number per unit formed mass is $1.3 \cdot 10^{-4} \Mo^{-1}$ for model $\mga$ and $8.8 \cdot 10^{-5} \Mo^{-1}$ for model $\maa$. Relaxing the condition of $M_1+M_2 < 1.4$ to all masses, $2.3 \cdot 10^{-4} \Mo^{-1}$ for model $\mga$ and $1.9\cdot 10^{-4} \Mo^{-1}$ for model $\maa$. As in the scenario proposed by \citet{Pak10}, when a scenario is biased to merging systems of high-mass ratio, the relative contribution from this scenario in the $\mga$ model is higher than the $\maa$ model.

\section{Conclusion and discussion}
\label{sec:con}
We studied the population of SNIa progenitors from merging double CO WDs with a combined mass exceeding the Chandrasekhar mass, the so-called DD progenitors. We considered two prescriptions of the CE phase. The CE evolution is a crucial ingredient in the formation of close double degenerate compact objects, but the process itself is still poorly understood. The first model assumes the $\alpha$-formalism for all CE. 
The second model is a combination of the $\alpha$-formalism and the $\gamma$-formalism (see Sect. \ref{sec:ce}). Typically, the first CE is described by the $\gamma$-scenario and the second by the $\alpha$-formalism, if mass transfer is unstable. 

We applied the updated version of the population synthesis code SeBa to simulate the population of DWDs and SNIa progenitors. At present, close DWDs (of all WD types) are the closest related systems to the DD SNIa progenitors that are visible in bulk. The mass ratio distribution of the DWDs in model $\maa$ is inconsistent with the observations. Using model $\mga$ the simulated population of DWDs compares well with observations, nevertheless, this is what the $\gamma$-formalism was designed to do. 

Recently, \citet{Web08} and \citet{Zor10} claimed that the predictive power of the $\gamma$-scenario is 
more restricted. They suggested that the $\alpha$-scenario is valid  
when sources of an energy other than the binding energy of the envelope is available, such as, the energy released by recombination in the common envelope. 
This could explain the high value of $\alpha$ found by \citet{Nel00} for the second CE, but certainly does not solve the problem for the first CE for which \citet{Nel00} found a value of $\alpha < 0$. 

The delay time distributions from our two models show the characteristic shape of a strong decay with time. This strong decay is expected when the delay time is dominated by the gravitational wave timescale ($t_{gr} \propto a^4$) and the distribution of orbital separations at DWD formation is similar to the initial (ZAMS) distribution of $N(a) da \propto a^{-1} da$ \citep{Abt83}.
The DTD from model $\mga$ fits the shape of the observed DTD best. 
\citet{Men10} also showed a DTD using the $\gamma$-scenario for the CE phase. They found that the DD DTD lies almost an order of magnitude lower in absolute rate than when using the $\alpha$-scenario. However, they used the $\gamma$-formalism for all CE phases. In our prescription (see Sect. \ref{sec:ce}) the $\gamma$-formalism is typically used in the first CE phase only. The reason for this is that in equal mass systems there is more angular momentum compared to unequal mass systems with similar orbits. \citet{Men10} and also \citet{Yun00, Rui09} and \citet{Cla11} showed DD DTDs using the $\alpha$-scenario (however their CE-prescriptions may differ slightly from Eq.\ref{eq:Egr}). Surprisingly, but as realised before, even though different groups used different binary evolution codes with different versions of the $\alpha$-CE and CE efficiencies, the DTDs of the DD channel are very uniform in that they show a strong decline with time \citep[see for example][for an overview]{Nel12}. 

Usually in synthetic DTD studies, the shape and normalisation of the DTD are discussed separately. This might not be valid any more, as more and more observed rates are available and the conversion from observational units to synthetic units (e.g. the star formation history (SFH) and rate in per K-band luminosity instead of per \Msolar\ of created stars) is better understood. For example, the SFH is often convolved with the DTD to estimate the SNIa rate in spiral galaxies like our Milky Way. The problem with this is that different assumptions for the Galactic SFH can significantly alter the theoretical Galactic SNIa rate. 
Since the SNIa rate follows the SFH with typical delay times of a few Gyr, the synthetic Galactic SNIa rate is very sensitive to the assumed SFH at recent times. 
When a constant SFH (of $\gtrsim 3\Mo\peryr$) is assumed, the SNIa rate is artificially enhanced compared with detailed SFHs that show a peak in the star formation at a few Gyr and a decline to $1\Mo\peryr$ at recent times, see e.g. \citet{Nel04}. 
In the observed SNIa rates of \citet{Mao10b} and \citet{Mao11} the detailed SFH of every individual galaxy or galaxy subunit was taken into account to reconstruct the DTD. Therefore it is no longer necessary to convolve the theoretical SNIa rate from a burst of star formation with an approximate SFH. The theoretical calculations of the SNIa rate from a single starburst can directly be compared with observations.   

We found that the normalisation of the DTD of model $\maa$ and $\mga$ do not differ much, even though the CE evolution is very different. The time-integrated number of SNIa in model $\maa$ ($3.3 \cdot 10^{-4} \Mo^{-1}$) is 70\% larger as in model $\mga$ ($2.0\cdot 10^{-4} \Mo^{-1}$). But most importantly, the simulated time-integrated numbers do not match the observed number of $2.3\pm 0.6 \cdot 10^{-3} \Mo^{-1}$ by \citep{Mao11} by a factor of $\sim 7-12$. 
If our understanding of binary evolution and initial parameter distributions is correct, the standard DD channel is not a major contributor to the SNIa rate.

For the SNIa model proposed by \citet{Pak10}, in which carbon burning is ignited in the merger process of two massive white dwarfs of nearly equal mass, we found an SNIa rate of $2.3 \cdot 10^{-5} \Mo^{-1}$ for model $\mga$ and $1.8 \cdot 10^{-5} \Mo^{-1}$ for model $\maa$. \citet{Pak10} founds that these systems resemble subluminous SNIa such as SN 1991bg. Assuming the fraction of 1991bg-like events to all SNIa events is 15$\pm$6\% \citep{Li01, Li11}, the observed event rate is $3.5 \pm 1.6 \cdot 10^{-4} \Mo^{-1}$. 
\citet{Ker10} proposed a model in which sub-Chandrasekhar WDs can explode as an SNIa. In this scenario two white dwarfs of nearly equal mass merge, though carbon ignition occurs only after the merger when the thick disk surrounding the remnant is accreted onto it. The event rate is $2.3 \cdot 10^{-4} \Mo^{-1}$ for model $\mga$ and $1.9 \cdot 10^{-4} \Mo^{-1}$ for model $\maa$. When only taking into account systems with a combined mass below 1.4$\Mo$, the rates are $1.3 \cdot 10^{-4} \Mo^{-1}$ and $8.8 \cdot 10^{-5} \Mo^{-1}$, respectively. In the scenario proposed by \citet{Pak10} and in the scenario by \citet{Ker10}, systems are required to have high mass ratios. We showed that the mass ratio distribution of DWDs depends on the prescription for the CE. When the $\gamma$-scenario is used, the average mass ratio of DWDs lies closer to one, which increases the SNIa rate in the above described scenario with respect to the $\alpha$-scenario. The rates of the channel proposed by \citet{Ker10} for systems with sub-Chandrasekhar masses are on the same order of magnitude as the rates of the standard DWD channel. Therefore the combination of the two models is not sufficient to explain the observed rates. 
For our synthetic rates of the DD scenario to match the observed SNIa rates, within our current model of binary evolution, the parameter space of the DD progenitors has to be increased severely, e.g. to include all CO-CO and CO-He mergers, which seems unlikely.

Alternatively (and if contributions from channels other than the DD are minor), our model underpredicts the fraction of standard DD SNIa progenitors in the entire DWD population. 
Our model of the visible population of DWDs predicts 0.9-2.9\% of the visible DWDs (depending on the model) to be SNIa progenitors. To match the observed rate of \citet{Mao11}, 10-30\% (excluding any errors on the observed and synthetic values) of the observed DWDs should lie in the SNIa progenitor region (upper left corner of Fig. \ref{fig:pop_Mt}). With 46 observed DWDs so far, 4-15 SNIa progenitors are expected without taking non-uniform selection effects into account. So far, only two systems have been found that possibly are SNIa progenitors, which makes it improbable, but not impossible, that our model underpredicts the number of DD SNIa progenitors. When the population of observed DWDs is increased, the fraction of SNIa progenitors amongst DWDs will give more insight into the validity of our knowledge of binary evolution of massive DWDs.

Concluding, although the shape of the DD DTD fits the observed DTD beautifully, the normalisation does not.  
An important point is that we did not optimise our model to fit the observed DTD in shape or number. 
We showed that the normalisation can be influenced by the metallicity; $\sim 30-60\%$ depending on the model for $Z=0.001$ with respect to $Z=0.02$. Furthermore, the normalisation depends on the initial mass function, the percentage of single stars, and the initial distribution of mass ratios and orbital periods. In this paper and in \citet{Nel12} we assumed the percentage of single stars to be 50\%. Results from e.g. \citet{Kou07} and \citet{Rag10} showed that the binary fraction might be as high as 70\% or more for A- and B-type stars, potentially raising the synthetic SNIa rate by a factor $<2$. Preliminary results show that the initial distribution of mass ratio and orbital separation affects the slope of the DTD, still the strong decline with time remains. Moreover, the integrated rates are not affected by factors sufficient to match the observed rate. 
Additional research is needed to study if the normalisation can be raised sufficiently to match the observed rate. 
If not, the main contribution to the SNIa rate comes from other channels, such as the SD scenario (e.g. supersoft sources), double detonating sub-Chandrasekhar accretors \citep[see e.g.][]{Kro10}, or Kozai oscillations in triple systems (\citealt{Sha12}; Hamers et al. in prep.). 

\begin{acknowledgements}
We thank Haili Hu, Marc van der Sluys and Lev Yungelson for providing results from their detailed stellar and binary evolution models. We also thank Dan Maoz for providing us his data of the observed SNIa rate and Frank Verbunt for his software to make the Roche lobe plots. This work was supported by the Netherlands Research Council NWO (grants VIDI [\# 639.042.813], VICI [\# 639.073.803], AMUSE [\# 614.061.608] and Little Green Machine) and by the Netherlands Research School for Astronomy (NOVA). 
\end{acknowledgements}

\bibliographystyle{aa}
\bibliography{bibtex_silvia_toonen}

\begin{appendix} 
\section{Most important changes to the population synthesis code \SeBa}
\label{appendix:}
\subsection{Treatment of wind mass-loss}
\label{appendix:wind}
Each star may lose mass in the form of a stellar wind. In a binary system the stellar wind matter from a binary component can be accreted by the companion star or lost from the system (see Appendix \ref{appendix:accretion}). This influences the binary parameters via the loss of mass and angular momentum from the system. We assume that the matter that is lost from the system carries a specific angular momentum equal to that of the star from which it originates. Furthermore, we assume that wind accretion onto the binary companion is Bondi-Hoyle accretion \citep{Bon44}, as re-formulated by \citet{Liv84}. 
The wind mass loss prescriptions for different types of stars used in \SeBa\ are updated e.g. to include metallicity dependency where possible. The prescriptions correspond to some degree to the recommendations by \citet{Hur00}. If multiple mass loss predictions are applicable to a star, we take the one that predicts the maximum mass loss rate. 

\begin{itemize}

\item For all types of luminous stars ($L>4000L_{\odot}$) from the main sequence (MS) to the asymptotic giant branch (AGB) we apply the empirical mass loss rate by \cite{Nie90} given by 

\begin{equation}
\dot M = 9.631 \cdot 10^{-15}\ R^{0.81}\ L^{1.24}\ M^{0.16}\ \Mo\ \peryr, 
\label{eq:wind:nieuw}
\end{equation}
where $\dot M$ is the mass accretion rate, $R$ the stellar radius in \Rsolar, $L$ the luminosity in $L_{\odot}$ and $M$ the stellar mass in \Msolar. We assume that the formalism of \cite{Nie90} is dependent on the initial metallicity as $\dot M(z) = (z/z_{\odot})^{1/2}\ \dot M(z_{\odot})$ \citep[see][]{Kud87}. 

\item For a massive MS star we give preference to the rates of \cite{Vin00, Vin01}. Where they do not apply, the rates of \cite{Nie90} are used. Massive MS suffer from strong winds driven by radiation pressure in lines and in the continuum. \cite{Vin00, Vin01} take into account multiple scattering effects of photons. They find good agreement between observations and theoretical mass-loss rates.

\item For stars in giant phases we adopt the empirical relation found by \cite{Rei75}, 
\begin{equation}
\dot M = 4\cdot 10^{-13}\  \frac{\eta RL}{M}\ \Mo\ \peryr.
\label{eq:wind:reimers}
\end{equation}
We assume a numerical prefactor of $\eta = 0.5$, see \cite{Mae89} and \cite{Hur00}. 

\item AGB stars can experience severe mass-loss caused by radiation pressure on dust that condensates in the upper atmosphere of the stars. Empirically, the mass-loss rate has been coupled to the period of large-amplitude radial pulsations $P_{\rm puls}$ \citep{Vas93}: 
\begin{equation}
{\rm log}\ P_{\rm puls}\ ({\rm days}) =  -2.07 + 1.94 \cdot {\rm log}(R) - 0.9 \cdot {\rm  log }(M).
\label{eq:wind:vas_p}
\end{equation}
We apply mass-loss to the envelope according to the prescription of \cite{Vas93}. During the superwind phase the radiation pressure driven wind is modelled by

\begin{equation}
\dot M = \frac{L}{cv_{\rm exp}},
\label{eq:wind:vas_mdot1}
\end{equation}
where $c$ represents the speed of light and $v_{\rm exp}$ the stellar wind expansion velocity. The latter is given by:

\begin{equation}
v_{\rm exp} ({\rm km\ s}^{-1}) = -13.5 +0.056 P_{\rm puls} ({\rm days}). 
\end{equation}
Furthermore, $v_{\rm exp}$ is constrained to the range 3.0-15.0 km $\persec$. 

Before the superwind phase, the mass loss rate increases exponentially with $P_{\rm puls}$ as
\begin{eqnarray}
\label{eq:wind:vas_mdot2}
{\rm log} \dot M\ (\Mo\ \peryr) = \quad \quad \quad \quad \quad \quad \quad \quad \quad \quad \quad \quad \quad \quad \\
\left\{
\begin{array}{lr}
 -11.4 + 0.0123 \cdot P_{\rm puls} & \text{if } M \leq 2.5,\\
 -11.4 + 0.0125 \cdot (P_{\rm puls}-100 \cdot (M-2.5))  & \text{if }  M > 2.5. \nonumber
\end{array} \right.
\end{eqnarray}

The mass loss rate of \citet{Vas93} is given by the minimum of Eq. \ref{eq:wind:vas_mdot1} and \ref{eq:wind:vas_mdot2}.

\item Luminous blue variables (LBVs) are extremely massive and luminous stars near the Humphreys-Davidson limit \citep{Hum94} with enormous mass-loss rates. We use the LBV mass loss prescription and implementation suggested by \citet{Hur00}:
\begin{equation}
\dot M = 0.1 \times \left (10^{-5}RL^{1/2}-1.0 \right )^3\ \left (\frac{L}{6.0\cdot 10^5}-1.0 \right )\ \Mo\ \peryr,
\label{eq:wind:lbv}
\end{equation}
if $L > 6.0\cdot 10^5L_{\odot}$ and $10^{-5}RL^{1/2} > 1.0$.

\item Wolf-Rayet stars are stars in a stage of evolution following the LBV phase where weak or no hydrogen lines are observed in their spectra. Like \cite{Hur00} we include a Wolf-Rayet-like mass-loss for stars with a small hydrogen-envelope mass ($\mu<1.0$ from their Eq. 97). The prescription itself, however, is different. We model it according to \cite{Nel01b}:
\begin{equation}
\dot M = 1.38\cdot 10^{-8}\ M^{2.87}\ \Mo\ \peryr. 
\label{eq:wind:wr}
\end{equation}
This is a fit to observed mass-loss rates from \cite{Nug00}. We multiply with a factor $(1-\mu)$ to smoothly switch on mass loss.

\item In addition to the evolution of ordinary hydrogen rich stars, the evolution of helium burning stars with hydrogen poor envelopes is simulated as well. For helium main-sequence stars with a mass $M > 2.5$\Msolar\, we assume the same relation as for Wolf-Rayet-like stars. For helium giants, either on the equivalent of the Hertzsprung or giant branch, we describe mass loss in a very general way similar to \cite{Nel01}. We presume 30\% of the mass of the envelope $M_{\rm env}$ will be lost during the naked helium giant phase with a rate that increases in time according to 
\begin{equation}
\Delta M_{\rm wind} = 0.3 M_{\rm env} \bigl[ \bigl(\frac{t+\Delta t}{t_{\rm f}}\bigr)^{6.8}-\bigl(\frac{t}{t_{\rm f}}\bigr) ^{6.8}\bigr],
\label{eq:Mwindheg}
\end{equation}
where $\Delta M_{\rm wind}$ is the amount of mass lost in the wind in \Msolar in a timestep $\Delta t$, $t_{\rm f}$ is the duration of the helium giant phase and $t$ the time since the beginning of the phase. 
\end{itemize}

Special attention has been given to prevent large wind mass losses in single timesteps because the mass loss prescriptions are very dependent on the stellar parameters of that timestep. For this reason we implemented an adaptive timestep in situations where strong winds are expected, e.g. at the tip of the giant branch. This procedure is accurate to differences in stellar mass of less than 4\% for masses below 12\Msolar.

\subsection{Accretion onto stellar objects}
\label{appendix:accretion}

Roche lobe overflow mass transfer and subsequent accretion can substantially alter the stars and the binary orbit. Mass accretion can affect the structure of the receiver star and its subsequent evolution. When more mass is transferred than the accretor can accrete, we assume that the non-accreted matter leaves the system with an angular momentum of 2.5 times the specific angular momentum of the binary \citep{Por96, Nel01}. For compact accretors we assume the matter leaves the system with the specific angular momentum of the compact remnant. In this section we discuss the limiting accretion rate, the response of the accretor to regain equilibrium, and the subsequent evolution of the new object for different types of accretors.

\subsubsection{Accretion onto ordinary stars}
For ordinary stars from MS to AGB stars, we distinguish between two types of accretion; accretion from a hydrogen-rich or a helium-rich envelope. Hydrogen-rich accretion can occur for example when a donor star ascends the giant branch and fills its Roche lobe. After it loses its hydrogen envelope, it can become a helium-burning core. When this helium star ascends the helium equivalent of the giant branch, a fraction of the helium-rich envelope can be transferred onto the accretor. We name this type of accretion 'helium accretion'. We assume that the accreted helium settles and sinks to the core instantaneously. The helium accretion rate is limited by the Eddington limit. Hydrogen is accreted onto the envelope of the receiver star. The accretion rate is bounded by the star's thermal timescale times a factor that is dependent of the ratio of Roche lobe radius of the receivers to its effective radius, as described by \citet{Por96}. The formalism is proposed by \citet{Pol94}, which is based on \citet{Kip77, Neo77} and \citet{Pac79}. If the mass transfer rate is higher than the maximum mass accretion rate, the excess material is assumed to leave the binary system.  

Because of the accretion, the star falls temporarily out of thermal equilibrium. While regaining equilibrium, the gas envelope surrounding the core puffs outward. Because we do not solve the equations of stellar structure and the stellar evolution tracks describe single stars in equilibrium, we add a procedure to account for a temporal increase in radius as in \citet{Por96}. This is important for example to determine if an accretor star fills its Roche lobe. It also affects the magnetic braking process and the Darwin-Riemann instability through the increased stellar angular momentum.
Note that the mass transfer rate is not dependent on the stellar radius in our simulations, so that the binary evolution is not critically dependent on out-of-equilibrium parameter values. 

Accretion can also affect the structure of the receiver star and its subsequent evolution. It is modelled by changing the stellar track and moving along the track. The former is described by the track mass, which is equivalent to the zero-age main-sequence mass that the star would have had without interaction. The latter is described by the relative age $t_{\rm rel}$ of the star. We distinguish two cases: 
\begin{itemize}
\item Rejuvenation of an MS star\\
Accretion onto an MS star rejuvenates the star. The star evolves similarly to a younger star of its new mass and its MS lifetime can be extended. It would show up in a Hertzsprung-Russell diagram as a blue straggler. For hydrogen accretion the track mass is always updated and the renewed relative age of the star

\begin{equation}
t'_{\rm rel} = t_{\rm rel}\ \frac{t'_{\rm ms}}{t_{\rm ms}} \frac{M}{M'},
\label{eq:rejuv_ms_h}
\end{equation}
where primes denote quantities after a small amount of mass accretion, $t_{\rm ms}$ the main-sequence lifetime, and $M$ the mass of the star. 

For helium accretion we assume the mass accretes to the core instantaneously and the track mass is increased accordingly. These stars appear older than for hydrogen accretion because more hydrogen has been burned previously. The rejuvenation process is described by

\begin{equation}
t'_{\rm rel} = t_{\rm rel}\ \frac{t'_{\rm ms}}{t_{\rm ms}} \frac{M}{M'} + \frac{\delta M t'_{\rm ms}}{0.1M'},
\label{eq:rejuv_ms_he}
\end{equation}
where we assume 10\% of the mass of the star will be burned during the MS phase.

\item Rejuvenation of a giant\\
During the giant phases the envelope is discoupled from the core in terms of stellar structure. The evolution of the star will therefore not be influenced directly by small amounts of hydrogen accretion to the envelope. The track is only updated when the new mass is larger than the track mass to account for severe hydrogen accretion. The mass before accretion can be much lower than the track mass because of wind mass loss, which can be strong for giants. For helium accretion to the core, the track is always updated. An exception to this is the early AGB where the helium core does not grow. In this stage there is a one-to-one relation between the helium core mass and the track mass \cite[Eq.66 in][]{Hur00}. 

When a giant accretor star moves to a new evolutionary track, we need to determine the location of the star along this track. In a more physical picture this means determining the relative age of the star $t_{\rm rel}$. For a giant its evolution is mainly determined by its core. Therefore for a given evolutionary track and core mass, the relative age is effectively constrained. For both types of accretion, we insist that the star stays in its same evolutionary state after its mass increase. When no solution can be found for $t_{\rm rel}$, the relative age is set to the beginning or end of the current evolutionary state and the track mass is varied to find a fitting track that ensures mass conservation. 

\end{itemize}

\subsubsection{Accretion onto helium-burning cores}
For accretion onto helium-burning stars that have lost their hydrogen envelopes, accretion is limited by the Eddington limit. Helium accretion onto a helium main-sequence star is similar as hydrogen accretion onto normal main-sequences stars. We assume that the star evolves similarly to a younger star of its new mass according to Eq. \ref{eq:rejuv_ms_h} where $t_{\rm ms}$ should be replaced by the helium main-sequence life time. 
We assume that for helium giants the envelope is discoupled from the core in terms of stellar structure, as with hydrogen-rich giants. Therefore we assume that the evolution of the giant is not affected and only update the track when the new mass is larger than the track mass.

The effect of hydrogen accretion onto helium stars is more complicated. If the hydrogen layer is sufficiently thick, the layer can ignite. This can significantly increase the radius of the star and essentially turn it into a born-again star on the horizontal or asymptotic giant branch. We studied the effect of hydrogen accretion to helium stars with stellar models simulated by the stellar evolution code STARS. This code models stellar structure and evolution in detail by solving the stellar structure equations. The code is based on \citet{Egg71} and includes updated input physics as described in \citet{Hu10}. The models do not include atomic diffusion. For mass accretion rates at ten percent of the Eddington rate of the accretor for $\sim 10^4-10^5 \yr$, the accreted hydrogen layer ignites. 
Helium stars that are more massive than $\sim 0.55M_{\odot}$ resemble horizontal branch stars after accretion, but most of the luminosity still comes from helium burning. For lower mass helium stars this is not the case, because the corresponding horizontal branch stars ($<3.5M_{\odot}$) can have ignited helium in a degenerate core, which strongly affects the characteristics of the star. For both mass ranges, the accretor expands by a factor $\sim 10-100$ compared to the original helium star. 
Because hydrogen accretion to helium stars is not very likely, we model this very simply. When more than 5\% of the total mass is accreted, the radius of the star is increased by a factor 50. With few exceptions, this leads to a merger of the two components. 

The effect of hydrogen accretion to helium giants is not known very well and additional research is necessary. For now, because it is very unlikely to happen, we treat it in the same way as helium accretion onto the envelope of the giant.

\subsubsection{Accretion onto remnants}
White dwarf, neutron star and black hole accretors can accrete with a maximum rate of the Eddington limit. 
If more mass is transferred, the surplus material leaves the system with the specific angular momentum of the compact remnant. 
For neutron stars and black holes we assume that the transferred mass is temporarily stored in a disk. From this disk, mass will flow onto the surface of the remnant with ten percent of the Eddington limit. We assume that a neutron star collapses onto a black hole when its mass exceeds $1.5\Mo$. 

For white dwarfs, the accretion process is more complicated because of possible thermonuclear runaways in the accreted material on the surface of the white dwarf. In \SeBa\ there are several options to model the effectiveness of the white dwarf to retain the transferred material. For hydrogen accretion we can choose between the efficiencies of \citet{Hac08} and \citet{Pri95}. For helium retention, the option is between \citet{Kat99} \citep[with updates from][]{Hac99} and \citet{Ibe96}. 

\subsection{Stability of mass transfer}
\label{appendix:stability}
A semi-detached system can become unstable in two ways. In a mass transfer instability, the Roche-lobe-filling star expands faster than the Roche lobe itself on the relevant timescale. In the other case tidal interactions lead to an instability \citep{Dar1879}. 

\subsubsection{Tidal instability}
A tidal instability can take place in systems of extreme mass ratios. When there is insufficient orbital angular momentum $J_{\rm b}$ that can be transferred onto the mass-losing star, the star cannot stay in synchronous rotation. Tidal forces will cause the companion to spiral into the envelope of the donor star. The tidal instability occurs when the angular momentum of the star $J_{\star} > \frac{1}{3} J_{\rm b}$, where

\begin{equation}
J_{\rm b} = Mm \sqrt{\frac{Ga}{M + m}},
\label{eq:jb}
\end{equation}
where $J_{\rm b}$ is the orbital angular momentum of the circularised binary, $a$ is the orbital separation, $M$ the mass of the donor star and $m$ the mass of the accretor star. The angular momentum $J_{\star}$ of a star with radius $R$ is given by
\begin{equation}
J_{\star} = k^2MR^2\omega, 
\label{eq:jstar}
\end{equation}
where $k^2$ is the gyration radius described by \citet{Nel01} and $\omega$ is the angular velocity of the donor star, which is assumed to be synchronised with the orbit. It is given by $\omega = 2\pi / P_{\rm b}$, where $P_{\rm b}$ is the orbital period. We model the inspiral according to the standard $\alpha$-CE (see Sect. \ref{sec:ce}). Owing to the expulsion of the envelope, the binary may evolve to a more stable configuration or merge. If the mass-losing star is a main-sequence star, we assume that the instability always leads to a merger.

\subsubsection{Mass transfer instability}
The stability of mass transfer from Roche lobe overflow and its consequences on the binary depend on the response of the radius and the Roche lobe of the donor star to the imposed mass loss (e.g. \citealt{Web85, Hje87} (hereafter HW87); \citealt{Pol94, Sob97}). We distinguish four modes of mass transfer; on the dynamical, thermal, nuclear timescale of the donor or on the angular-momentum-loss timescale. The response of the accretor star to the mass that is transferred onto it and the effect of this on the orbit is described in Appendix \ref{appendix:accretion}. The response of the donor star to mass loss is to readjust its structure to recover hydrostatic and thermal equilibrium. 
The dynamical timescale to recover hydrostatic equilibrium is short compared to the thermal timescale. For mass transfer to be dynamically stable, the dynamical timescale of the star is important. 
The change in radius due to adiabatic adjustment of hydrostatic equilibrium is expressed as a logarithmic derivative of the radius with respect of mass,
\begin{equation}
\zeta_{\rm ad} = \left ( \frac{d\ \rm ln\ R}{d\ \rm ln\ M} \right )_{\rm ad},
\end{equation}
where $M$ and $R$ are the mass and radius of the donor star. The assumed values of $\zeta_{\rm ad}$ are shown in Table \ref{tbl:zeta}.

The response of the Roche lobe $R_{\rm L}$ of the donor star is expressed as the logarithmic derivative of the Roche lobe radius with respect to mass:

\begin{equation}
\zeta_{\rm L} = \frac{d\ \rm ln\ R_{\rm L}}{d\ \rm ln\ M}. 
\end{equation}
The value of $\zeta_{\rm L}$ is calculated numerically by transferring a test mass of 10$^{-5}$ \Msolar. 
Because $\zeta_{\rm L} = \zeta_{\rm L}(M_1, M_2, a)$, $\zeta_{\rm L}$ is dependent on the mass accretion efficiency of the secondary, and therefore on the mass accretion rate of the test mass. For instance, for high mass ratios $q \gg 1$ the loss of some mass and corresponding angular momentum can have a stabilising effect on the mass-transferring binary. To determine the dynamical stability of mass transfer, we assume that the mass transfer rate of the test mass is on the thermal timescale of the donor star:
 
\begin{equation}
\tau_{\rm th} = \frac{GM^2}{RL}.  
\end{equation}

\begin{enumerate}
\item When $\zeta_{\rm L} >\zeta_{\rm ad} $, mass transfer is dynamically unstable. We model this as a CE phase, as described in Sect. \ref{sec:ce}. \footnotemark
\end{enumerate}

\footnotetext{
When the timescale of the CE phase becomes relevant, we assume that it proceeds on a time scale $\tau$ given by the geometric mean of the thermal $\tau_{\rm th}$ and dynamical $\tau_{\rm d}$ timescales of the donor \citep[see][]{Pac72}:
\begin{equation} 
\tau = \sqrt{\tau_{\rm d} \tau_{\rm th}}; \quad \quad \tau_{\rm d} = \sqrt{\frac{R^3}{GM}}. 
\end{equation}
}

When $\zeta_{\rm L} <\zeta_{\rm ad}$ mass transfer is dynamically stable. The donor star is able to regain hydrostatic equilibrium and shrinks within its Roche lobe on a dynamical timescale. To determine if the donor star is also able to regain thermal equilibrium during mass transfer, we calculate the change in the radius of the star as it adjusts to the new thermal equilibrium:
\begin{equation}
\zeta_{\rm eq} = \left ( \frac{d\ \rm ln\ R}{d\ \rm ln\ M} \right )_{\rm th}.
\end{equation}
The assumed values for $\zeta_{\rm eq}$ are described in Table \ref{tbl:zeta}.

To calculate the response of the Roche lobe $\zeta_{\rm L,eq}$, we assume that the mass transfer rate of the test mass is on the nuclear evolution timescale:
\begin{equation}
\tau_{\rm nuc} = dt\ \frac{R}{dR_{\rm eq}},
\end{equation}
where $R$ represents the equilibrium radius of the star according to the single-star tracks. $dR_{\rm eq}$ is the change in $R$ in a short timestep $dt$ without binary interactions.

\begin{enumerate}
\setcounter{enumi}{1}

\item When $\zeta_{\rm L,eq} < min(\zeta_{\rm eq},\zeta_{\rm ad})$ mass transfer is driven by the expansion of the stellar radius due to its internal evolution. 

\item When $ \zeta_{\rm L,eq} > min(\zeta_{\rm eq},\zeta_{\rm ad}) $, the mass transfer is thermally unstable and proceeds on the thermal time scale of the donor.
\newline

\item The previous modes of mass transfer are caused by an expanding donor star. The final mode is caused by shrinking of the orbit caused by angular momentum loss. We assume that this mode takes place when the corresponding timescale $\tau_{\rm J}$ is shorter than the timescales at which the other three modes of mass transfer take place. Angular momentum loss can happen due to gravitational wave radiation $\dot J_{\rm gr}$ \citep{Kra62} and magnetic braking $\dot J_{\rm mb}$ \citep{Sch62, Hua66, Sku72, Ver81}. Mass transfer proceeds on the time scale on which these processes occur:
\begin{equation}
\tau_{\rm J} = \frac{J_{\rm b}}{\dot J_{\rm gr}+\dot J_{\rm mb}},  
\end{equation}
where $J_{\rm b}$ is the angular momentum of the circularized binary given by Eq. \ref{eq:jb}. Next we discuss the assumptions and implications of                                                                                                                                                                                              $\dot J_{\rm gr}$ and $\dot J_{\rm mb}$.

Gravitational wave radiation most strongly influences close binaries since it is a strong function of orbital separation. The change in orbital separation $\dot a$ averaged over a full orbit is given by \citep{Pet64}

\begin{equation}
\dot a_{\rm gr} = -\frac{64}{5}\ \frac{G^3 Mm (M+m)}{c^5a^3(1-e^2)^{7.2}}\ \left (1 + \frac{73}{24}e^2 + \frac{37}{96}e^4 \right ), 
\end{equation}
where $\dot J_{\rm gr}/J_{\rm b} = \dot a_{\rm gr}/(2a)$. 

Magnetic braking mostly affects low-mass stars within the mass range of $0.6 \lesssim M/M_{\odot} \lesssim 1.5 $. These stars suffer from winds that are magnetically coupled to the star. Although the mass loss in this process is negligible, the associated angular momentum loss can be severe \citep{Rap83}: 

\begin{equation}
\frac{\dot J_{\rm mb}}{J_{\rm b}} = -\frac{3.8\cdot 10^{-30}R_{\odot}^{4-\beta}(M+m)R^{\beta}\omega^2 }{ma^2},
\end{equation}
where $\beta$ is a parameter that represents the dependence of the braking on the radius of the donor star. We take $\beta = 2.5$. 

\end{enumerate}

\begin{table}
\caption{
Values of the adiabatic $\zeta_{\rm ad}$ and thermal $\zeta_{\rm eq}$ response of the radius to mass loss for different types of stars.}

\begin{tabular}{|c|l|c|c|}
%\hline \hline
\thickhline
 $k$ & Evolutionary state & $\zeta_{\rm ad}$ & $\zeta_{\rm eq}$ \\
%\hline
\thickhline
0,1& Main-sequence: & & \\
&$M_{\rm o} < 0.4$ & $-1/3$& 0\\
&$0.4< M_{\rm o} < 1.5$ & 2& 0.9\\
&$M_{\rm o} > 1.5$ & 4& 0.55\\

\hline
2 & Hertzsprung gap: &  & \\
&$M_{\rm o} < 0.4$ & 4 & 0 \\
&$M_{\rm o} > 0.4$ & 4 & -2 \\

\hline
3&First giant branch: & & \\
&$\cdot$ shallow convective layer & 4 & 0 \\
&$\cdot$ deep convective layer & HW87 & 0 \\

\hline
4 &Horizontal branch: &  & \\
&$M_{\rm o} < M_{\rm Hef}$ & 4 & 4 \\

&$M_{\rm Hef} < M_{\rm o} < M_{\rm FGB}$: &  &  \\
&$\cdot$ decent along GB & as $k=3$ & 0\\
&$\cdot$ blue phase & 4& 4 \\

&$M_{\rm o} > M_{\rm FGB}$: &  &  \\
&$\cdot$ blue phase & 4& -2\\
&$\cdot$ ascent to AGB & HW87&0 \\

\hline
5,6&Asymptotic giant branch & HW87 & 0\\

\hline
7&Helium main-sequence: &  & \\
&$M<0.2$ & 15 & -0.19 \\
&$M>0.2$ & 15 & 1 \\

\hline
8,9& Helium giant: &  & \\
&$M_{\rm c} < 0.4$ & HW87 & $-1/3$ \\
&$M_{\rm c} > 0.4$ & HW87 & -2 \\

\hline
10,11,12& White dwarf & $-1/3$& $-1/3$ \\

\hline
\end{tabular}
\begin{flushleft}
\tablefoot{The types of stars correspond to the definition by \citet{Hur00} expressed by their integer $k$. The stellar tracks are distinguished by the mass $M_{\rm o}$, which is equivalent to the ZAMS mass the star would have had without interaction. $M_{\rm Hef}$ and $M_{\rm FGB}$ are defined by Eq. 2 and 3 from \citet{Hur00}. $M_{\rm Hef}$ represents the maximum initial mass for which helium ignites degenerately in a helium flash, which is $\sim 2$\Msolar\ for solar metallicities. $M_{\rm FGB}$ is the maximum initial mass for which helium ignites on the first giant branch, which is $\sim 13$\Msolar\ for solar metallicities. $M$ is the total mass of the star and $M_{\rm c}$ the mass of the core. HW87 represents \citet{Hje87}, who calculated $\zeta_{\rm ad}$ for condensed polytropes, consisting of a compact core surrounded by an envelope with polytropic index $n = 3/2$. For stars on the first giant branch there are two prescriptions of $\zeta_{\rm ad}$. If the convective zone in the upper layers of the envelope is shallow \citetext{fits from Yungelson, private communication}, we assumed the envelope responds to mass loss in a similar manner as radiative envelopes. }
\end{flushleft}
\label{tbl:zeta}
\end{table}

\end{appendix}
\end{document}